\documentstyle[11pt,aaspp4]{article}

\def\distrbn{distribution }
\def\gal{galaxy }

\def\be{\begin{equation}}
\def\ee{\end{equation}}
\def\eqref#1{(\ref{eqn:#1})}

\def\overdeng{\rho_g/{\overline \rho}_g}
\def\overdenm{\rho_m/{\overline \rho}_m}
\def\rhobar{{\overline \rho}}
\def\hmpc{{h^{-1}\;{\rm Mpc}}}
\def\hvol{{h^3\;{\rm Mpc}^{-3}}}
\def\hkpc{{h^{-1}\;{\rm kpc}}}
\def\kms{{\rm \;km\;s^{-1}}}
\def\hubunits{\kms\;{\rm Mpc}^{-1}}

\def\md{morphology-density }
\def\fsp{F_{\rm Sp}}
\def\fs0{F_{\rm S0}}
\def\fe{F_{\rm E}}
\def\divv{\nabla\cdot{\bf v}}
\def\rec{reconstruction }
\def\gau{Gaussianization }
\def\pscz{PSCz }
\def\zel{Zel'dovich }

\begin{document}

\title{RECONSTRUCTION ANALYSIS OF THE IRAS POINT SOURCE CATALOG REDSHIFT SURVEY}

\author{
Vijay K. Narayanan\altaffilmark{1},
David H. Weinberg\altaffilmark{1},
E. Branchini\altaffilmark{2}, 
C. S. Frenk\altaffilmark{3},
S. Maddox\altaffilmark{4},
S. Oliver\altaffilmark{5},
M. Rowan-Robinson\altaffilmark{5}, and
W. Saunders\altaffilmark{6}
}
\altaffiltext{1}{Department of Astronomy, The Ohio State University, Columbus, OH 43210, U.S.A; Email: vijay,dhw@astronomy.ohio-state.edu} 

\altaffiltext{2}{Kapteyn Sterrewacht, Rijksuniversiteit Groningen, Postbus 800, 9700, AV Groningen, The Netherlands; Email: branchin@astro.rug.nl}

\altaffiltext{3}{Department of Physics, University of Durham, Science Laboratories, South Rd, Durham, DH1, SLE, U.K; Email: C.S.Frenk@durham.ac.uk}

\altaffiltext{4}{Department of Physics, University of Nottingham, Nottingham, U.K; Email: ppzsjm@unix.ccc.nottingham.ac.uk}

\altaffiltext{5}{Imperial College of Science, Technology, and Medicine, Blackell Laboratory, Prince Consort Road, London,  SW1 2EZ, U.K; Email: s.oliver,m.rrobinson@ic.ac.uk}

\altaffiltext{6}{Institute for Astronomy, University of Edinburgh, Blackford Hill, Edinburgh EH9 3JS, U.K; Email: will@roe.ac.uk}

\begin{abstract}
We present the results of \rec analysis of the galaxy distribution 
in a spherical region of radius $50 \hmpc$ centered on the Local Group,
as mapped by the IRAS Point Source Catalog Redshift Survey (PSCz).
We reconstruct this galaxy distribution  using 15 different models for 
structure formation in the universe, each model consisting of a set of 
assumptions regarding the value of the cosmological mass density parameter 
$\Omega_{m}$, and the amplitude and nature of the biasing between IRAS 
galaxies and the underlying mass.
For every model, we also reconstruct ten mock \pscz catalogs derived
from the outputs of numerical simulations that have the appropriate values of
$\Omega_{m}$ and bias.
We quantify the accuracy of a reconstruction using a variety of statistics
and compare the accuracy of each reconstruction of the \pscz catalog to the
accuracy expected based on the mock catalog reconstructions of the 
corresponding model.
We find that gravitational instability of Gaussian primordial mass 
density fluctuations can account for the galaxy distribution 
in the \pscz catalog, at least for some plausible assumptions about the 
value of $\Omega_{m}$ and the biasing between IRAS galaxies and mass.
However, unbiased models in which IRAS galaxies trace mass fail to reconstruct 
the \pscz catalog accurately, both for $\Omega_{m}=0.4$ and for $\Omega_{m}=1$.
Low $\Omega_{m}$ models in which IRAS galaxies are antibiased with respect
to the mass distribution are the most successful in reconstructing the
\pscz catalog.
In particular, a model with $\Omega_{m}=0.4$ and IRAS galaxies related 
to the mass distribution according to the predictions of a semi-analytic
galaxy formation model is very successful in reproducing the properties
of the \pscz galaxy distribution.
\end{abstract}


\section{Introduction}

Understanding the formation and evolution of large scale structure (hereafter 
LSS) in the universe is one of the foremost problems in cosmology.
In the standard approach to studying LSS, one starts with a model
for the primordial mass density fluctuations, uses analytical approximations
or numerical simulations to predict the ensemble averaged statistical 
properties of galaxy clustering, and compares them with those of the 
observed galaxy distribution, assuming that we observe fair sample
of the universe.
However, we cannot expect a simulation started from random initial 
conditions to reproduce the specific structures observed in a galaxy redshift 
catalog, even if the {\it statistical} properties of the galaxy clustering 
are correct.
Reconstruction analysis is a complementary approach to the study of LSS in 
which one works backward from the observed galaxy distribution
to the initial mass density fluctuations in the same region of space, 
then evolves these model initial conditions forward in time to the present day
using an N-body code.
The reconstruction method incorporates assumptions about the properties
of primordial fluctuations, the values of cosmological parameters, and the 
``bias'' between the galaxy and mass distributions.
These assumptions can be tested by comparing in detail the evolved
reconstruction to the original galaxy redshift data.
A by-product of the \rec analysis is a detailed model for the origin
and evolution of familiar, well studied structures in the local 
universe, such as the Great Attractor, the Perseus-Pisces supercluster, and
the Sculptor Void.

In this paper, we present the results of reconstruction analysis of the IRAS
Point Source Catalog Redshift survey (hereafter PSCz, Saunders et al. 1995; 
Canavezes et al. 1998),  using the ``hybrid'' \rec method described
by Narayanan \& Weinberg (1998, hereafter NW98).
First, we create the galaxy density field smoothed with a 
Gaussian filter of radius $R_{s} = 4 \hmpc$ 
(where $h \equiv H_0 / 100\;\hubunits$), correct for the effects of bias and 
redshift-space distortions, and derive the smoothed initial mass density field
by tracing the evolution of these density fluctuations backward in time.
We then evolve these initial density fluctuations forward using an 
N-body code, assuming a value for $\Omega_{m}$, and compare in detail 
the clustering properties of the reconstructed PSCz galaxy distribution to 
those of the input \pscz galaxy distribution, either assuming that 
galaxies trace mass or selecting galaxies from the N-body particle 
distribution using a biasing prescription.
For our purposes, therefore, a model of structure formation consists of
a value of $\Omega_{m}$, a bias factor $b$ that gives the ratio of rms
galaxy and mass fluctuations on a scale of $8 \hmpc$, and an explicit
biasing scheme that specifies how galaxies are to be selected from the
large scale mass distribution.
All of our models assume that structure grew by gravitational instability 
from Gaussian primordial fluctuations --- these are the implicit assumptions 
of the \rec method itself.
We reconstruct the \pscz catalog using 15 different models and quantify
the accuracy of each \rec using a variety of clustering statistics.
Even if the model assumptions are correct, we do not expect to reproduce
the observed structure exactly, because we begin with imperfect data and 
because the \rec method cannot invert gravitational evolution in the 
strongly non-linear regime.
For each statistic, we therefore rank the accuracy of the model's \pscz 
\rec with respect to the reconstructions of ten mock \pscz catalogs derived 
from the outputs of N-body simulations of the model under consideration.
Finally, we use these ranks to evaluate the success of the \pscz \rec
for each model, to constrain $\Omega_{m}$, and to test models of bias between
the mass and IRAS galaxy distributions in the real universe.

The hybrid \rec technique (NW98) that we use for our \pscz analysis
combines the complementary desirable features of the \gau mapping 
method (Weinberg 1992), which assumes Gaussian primordial fluctuations 
and a monotonic relation between the smoothed initial mass density field
and the smoothed final galaxy density field, and the dynamical \rec methods of
Nusser \& Dekel (1992) and Gramann (1993a), which are based on the momentum 
and mass conservation equations, respectively, under the approximation that 
the comoving trajectories of mass particles are straight lines 
(the Zel'dovich [1970] approximation).
In the hybrid method, we first recover the smoothed initial
density field from the smoothed final mass density field 
using a modified form of the dynamical methods, then Gaussianize this 
recovered {\it initial} density field to robustly recover the initial 
fluctuations even in the non-linear regions (an approach also used by
Kolatt et al. 1996).\footnote[1]{To ``Gaussianize'' a field, one
applies a local, monotonic mapping that enforces a Gaussian 1-point
PDF.}
For a \rec that incorporates biased galaxy formation, we precede the
dynamical \rec step with a  step that maps the smoothed galaxy density
field monotonically to a smoothed mass density field with the 
theoretically expected (non-linear, non-Gaussian) PDF.
The hybrid method is described and tested in detail in NW98, who show that
it can reconstruct a galaxy redshift survey more accurately than either the 
\gau method or the dynamical \rec methods alone.
Comparison of the hybrid method to a variety of alternative \rec schemes, 
including the Path Interchange \zel Approximation (PIZA) method of
Croft \& Gazta\~{n}aga (1997), is given by Narayanan \& Croft (1999).

Earlier attempts to reconstruct observational data include the reconstruction
of the Perseus-Pisces redshift survey of Giovanelli \& Haynes (1989)
by Weinberg (1989) using the Gaussianization technique and the reconstruction
of the IRAS $1.2$Jy redshift survey (\cite{fisher95}) by Kolatt et al. (1996)
using the dynamical scheme of Nusser \& Dekel (1992).
This dynamical scheme was also used by Nusser, Dekel \& Yahil (1995) to
recover the PDF of the initial density field from the IRAS $1.2$Jy
redshift survey.
The primary requirements for a redshift survey to be suitable for 
\rec analysis are: 
(a) good sky coverage and depth, so that the gravitational influence of 
regions outside the survey boundaries is small,
(b) dense sampling to reduce shot noise errors, and 
(c) a well understood selection function, to allow accurate construction
of the observed galaxy density field.
With respect to these criteria, \pscz is a substantial improvement on
samples used in previous analyses, and it is the best all-sky redshift survey
that exists today.
The PSCz is a redshift survey of all galaxies in the IRAS Point Source Catalog
whose flux at $60\ \mu m$ is greater than $0.6$Jy.
It contains about $15,500$ galaxies distributed over $84.1\%$ of the sky,
excluding only some regions of low Galactic latitude where the extinction
in the $V$ band as estimated from the IRAS $100\ \mu m$ background is greater 
than 1.5 mag. (mainly the low Galactic latitude zone 
$\vert b \vert < 5^{\circ}$), the Magellanic clouds, some odd patches 
contaminated by Galactic cirrus, and two strips in ecliptic longitude 
not surveyed by the IRAS satellite.

The plan of this paper is as follows.
In \S2, we describe the hybrid reconstruction method used to reconstruct 
galaxy redshift surveys, outline the assumptions involved in the 
\rec analysis, and list all the steps involved in reconstructing the 
\pscz catalog in the order in which they are implemented.
In \S3, we describe the various models that we use to reconstruct the
\pscz catalog, and our construction of the mock \pscz catalogs for each model
from the outputs of N-body simulations.
In \S4, we illustrate the results of reconstruction analysis for 6 of our 
15 models, using a variety of statistics.
We quantify the accuracy of the \pscz reconstruction of a model using 
a ``Figure-of-Merit'' for each statistic, and rank the \pscz \rec with 
respect to all the mock catalogs for that model.
We summarize the results of reconstructing the \pscz catalog for the full set
of 15 models in \S5, and describe the criteria for classifying a model as 
Accepted, or Rejected, based on its rankings.
We review our results and discuss their implications in \S6, drawing
general conclusions about the value  of $\Omega_{m}$, the amplitude of
mass density fluctuations, the bias between IRAS galaxies and mass, and the
viability of gravitational instability with Gaussian initial conditions
as an explanation for the structure observed in the \pscz redshift survey.
A brief overview of our results can be obtained from Figures 3, 7, and 13,
and Table 2.

\section{Reconstruction Analysis}

We reconstruct the galaxy distribution in the \pscz catalog using the
``hybrid'' \rec method of NW98.
We refer the reader to NW98 for a detailed discussion of the method,
including its general motivation and tests of its accuracy on N-body
simulations.
Here we provide a summary of the assumptions made in the \rec (\S2.1),
justification of our choice of smoothing length and sample radius for
\pscz (\S2.2), and a step-by-step description of the \rec procedure as
applied to PSCz (\S2.3).

\subsection{Assumptions}

Reconstruction analysis of a galaxy redshift catalog incorporates a 
number of assumptions, at various stages.
These include assumptions about the cosmological parameters, about the 
nature of the primordial mass density fluctuations, about the process
of structure formation, and about the physics of galaxy formation.
The assumptions in our analysis are:
\begin{description}
\item[{(1)}:] Structure formed by gravitational instability.
The reconstruction procedure traces the evolution of density
fluctuations backward in time under the assumption that the LSS formed from 
the gravitational instability of small amplitude fluctuations in the 
primordial mass density field.
This assumption is also implicit when the power-restored initial density
field is evolved forward in time using a gravitational N-body code.
\item[{(2)}:] The primordial density fluctuations form a Gaussian random 
field, as predicted by simple inflationary models for the origin of these 
fluctuations (\cite{guth82}; \cite{hawking82}; \cite{starobinsky82}; 
Bardeen, Steinhardt, \& Turner 1983).
This assumption is the basis of the \gau step of the \rec procedure.
\item[{(3)}:] The values of $\Omega_{m}$ and the bias factor 
$b \equiv \sigma_{8g}/\sigma_{8m}$, where $\sigma_{8m}\ (\sigma_{8g})$ is the rms
fluctuation of the mass (IRAS galaxy) distribution in spheres of radius 
$8 \hmpc$.
We vary these assumptions from one \rec to another.
We use the value of $\Omega_{m}$ in correcting for redshift-space distortions 
and in forward evolution of the reconstructed initial conditions.
We use the value of $\sigma_{8m} = \sigma_{8g}/b$ to normalize the 
forward evolution simulations.
Note that throughout the paper we use $b$ to refer to the rms fluctuation
bias on the $8 \hmpc$ scale.
\item[{(4)}:] The shape of the primordial power spectrum.
In contrast to the amplitude $\sigma_{8m}$, changes to the power spectrum shape
make little difference to our results, because the shape is used only to 
compute corrections to or extrapolations of the initial power spectrum
recovered from the observational data.
\item[{(5)}:] The evolved galaxy density field is a monotonic function of the 
evolved mass density field, once both are smoothed over scales of
a few $\hmpc$.
This assumption, together with the value of $\sigma_{8m}$ and the assumption
of Gaussian initial conditions, allows us to recover the smoothed mass density
field from the smoothed galaxy density field in preparation for the 
time-reversed dynamical evolution.
\item[{(6)}:] An explicit biasing scheme, i.e., a prescription for
relating the underlying mass distribution  to the observable galaxy 
distribution.
This scheme does not influence the recovery of initial conditions, but it is
needed to select galaxies from the N-body simulation evolved from these
initial conditions, and hence to compare the \rec to the input data.
Most of our biasing schemes have a single free parameter that controls the
strength of the bias.
We set the value of this parameter to obtain the desired bias factor 
$b$ (assumption 3).
\end{description}

\subsection{Choice of smoothing length and sample radius}

Since the \pscz is a flux-limited survey, the number density of galaxies 
in the catalog decreases with distance from the observer.
Consequently, the shot-noise in the PSCz galaxy distribution increases with 
distance.
However, the \gau procedure in the \rec analysis relies on the assumption 
that the rms amplitude of galaxy density fluctuations remains the same 
throughout the reconstruction volume and that the contribution to these 
fluctuations from shot-noise in the galaxy distribution is negligible.
In order to ensure that the shot-noise remains small and does not increase 
with distance from the observer, we create a volume-limited PSCz sub-catalog
in which the number density of galaxies remains constant throughout the
reconstruction volume.

Much of the diagnostic power of the reconstruction analysis stems from the 
fact that non-linear gravitational evolution transfers power from large 
scales to small scales (\cite{melott90}; \cite{beacom91}; 
Little, Weinberg, \& Park 1991; \cite{soda92}; \cite{bagla97}).
This power transfer erases the information about the initial mass fluctuations
on scales below the non-linear scale (Fourier wavenumbers $k \geq k_{\rm nl}$).
Consequently, a reconstruction method that recovers the initial fluctuations
accurately up to the non-linear scale $k = k_{\rm nl}$ can reproduce many 
features of the evolved structures even on smaller scales $(k > k_{\rm nl})$, 
though not, of course, the finer details of these features.
Since the rms fluctuation of the IRAS galaxy distribution in spheres of
radius $ 8 \hmpc$ is about 0.7 (Saunders, Rowan-Robinson \& Lawrence 1992;
\cite{fisher94}; \cite{moore94}), we need to recover the initial density 
field accurately at least up to this scale in order to take advantage of the 
power-transfer phenomenon.
We will therefore reconstruct the PSCz catalog using a Gaussian smoothing
length of $R_{s} = 4 \hmpc$, which corresponds to a top-hat smoothing scale of 
about $6.6 \hmpc$.

We create the volume-limited sub-catalog by selecting all the galaxies in the
PSCz located within a volume-limiting radius that are bright enough to
be included in the survey even when they are located at this volume-limiting
radius.
We choose this volume-limiting radius, $R_{1}$, based on a compromise between
two conflicting requirements.
First, the reconstruction volume should be large enough that it contains
many independent smoothing volumes.
This criterion pushes us to choose a large value for $R_{1}$.
Second, the shot-noise in the galaxy distribution of the volume-limited
catalog should be small and remain constant with distance from the observer.
This condition requires a uniformly high number density of galaxies, pushing
us to adopt a smaller volume-limiting radius.
Since we reconstruct the \pscz catalog using a Gaussian smoothing length of
$R_{s} = 4 \hmpc$, we fix $R_{1}$ so that the mean inter-galaxy separation
at $R_{1}$ is $\bar d \equiv n_{g}^{-1/3} = \sqrt{2}R_{s} = 5.6 \hmpc$.
We adopt this criterion $\bar d = \sqrt{2}R_{s}$ based on the rule of thumb
suggested by Weinberg, Gott, \& Melott (1987), to obtain the largest 
possible \rec volume while keeping shot-noise in the galaxy distribution 
small enough to have little effect on the smoothed galaxy density field.

We compute the number density of galaxies as a function  of the distance from 
the observer in the PSCz catalog  using the maximum-likelihood method described
by Springel \& White (1998).
We find that the number density of galaxies drops to 
$0.005 \ {\rm h^{3}Mpc^{-3}} = (5.6 \hmpc)^{-1/3}$ at a distance of 
$R_{1} = 50 \hmpc$ from the observer.
We then select all the galaxies in the PSCz catalog that are  bright
enough to be included in the survey even if they are placed at a distance
of $50 \hmpc$ from the Local Group.
The galaxies selected in this manner are then included in the \pscz 
sub-catalog, which is thus volume-limited to $R_{1} = 50 \hmpc$.
The luminosities at $60\ \mu m$ of the galaxies in this sub-catalog
satisfy the condition 
$\log_{10} \left( \frac{L_{60}}{L_{\odot}} \right) > 9.37$
(for $h = 1$).

\subsection{Step-by-step description}

The steps involved in reconstructing a $50 \hmpc$, volume-limited 
subset of the \pscz catalog for a given set of model assumptions are as
follows:
\newline
{\it Step 1: }
Create an all-sky galaxy distribution by ``cloning'' the galaxy
distribution to fill in the regions excluded in the PSCz survey.
The \pscz catalog does not include galaxies in  regions of low Galactic
latitude where there is substantial obscuration by dust.
However, this region could be dynamically important, since
the Perseus-Pisces supercluster and the Hydra-Centaurus supercluster are
both located close to the Galactic plane and could even extend across it.
Hence, we fill in the region with Galactic latitude 
$\vert b_{\rm cut} \vert < 8^{\circ}$, using the cloning technique introduced 
by Lynden-Bell, Lahav \& Burstein (1989; see also Yahil et al. 1991).
We divide this region into 36 angular bins of $10^{\circ}$ in longitude and 
divide the redshift range in each angular bin into bins of $1000 \kms$.
In each longitude-redshift bin, we assign $N(l,z)$ artificial galaxies, where 
$N(l,z)$ is equal to a random Poisson deviate whose expectation value is
equal to the average density of the corresponding longitude-redshift bins
in the adjacent strips 
$\vert b_{\rm cut} \vert < \vert b \vert < 2\vert b_{\rm cut} \vert$, times
the volume of the bin.
If there is a real \pscz galaxy in any of these longitude-redshift bins,
we include it in place of an artificial galaxy.
We fill the masked regions at high Galactic latitudes with a random 
distribution of artificial galaxies having the observed mean density.
The flux distribution of the artificial galaxies is identical to those of 
the real galaxies in the PSCz catalog.
We tested (using mock \pscz catalogs) alternate methods of handling 
the mask region, including assigning galaxies at random locations within the 
mask region at the mean galaxy density, as well as ignoring all the galaxies 
in the  mask region.
We found that the cloning technique always leads to the most accurate
reconstruction of the galaxy distribution.
However, the mask region does not influence \rec analysis as much
as it influences say, the analysis of the cosmological galaxy dipole
(Rowan-Robinson et al. in preparation).
\newline
{\it Step 2: }
Select a volume-limited galaxy distribution from the 
flux-limited PSCz catalog so that the shot-noise in the volume-limited 
catalog is small and remains constant throughout the \rec volume.
Based on the selection function of the PSCz survey, we choose the 
volume-limiting radius $R_{1} = 50 \hmpc$, where the mean inter-galaxy 
separation is $\bar d = \sqrt{2}R_{s} = 5.6 \hmpc$.
\newline
{\it Step 3: }
Compute the smoothed galaxy density field in redshift space.
We create the PSCz galaxy density field in redshift space by cloud-in-cell 
(CIC) binning (Hockney \& Eastwood 1981) the volume-limited galaxy distribution
onto a $100^{3}$ cubical grid that represents $200 \hmpc$ on a side.
The Local Group observer is at the center of this cube, and the three  sides
of the cube are oriented along the axes of the Supergalactic
coordinate system.
Since the dynamical component of the hybrid \rec method traces the evolution 
of the gravitational potential backward in time, it is necessary to model 
the gravitational field in the regions beyond the boundaries in order to 
reconstruct the density field accurately near the edges of the 
volume-limited catalog.
We therefore supplement the volume-limited density field with the density
field in an annular region $20 \hmpc$ thick beyond the volume-limiting 
radius of $50 \hmpc$.
We form the density field in this annular region by weighting each galaxy
by the inverse of the selection function of the flux-limited PSCz survey
at the location of the galaxy.
The full galaxy density field is therefore constructed from a volume-limited 
catalog in the region $0 < R < R_{1} = 50 \hmpc$ and from a flux-limited 
catalog in the region $R_{1} < R < R_{2} = 70 \hmpc$.
We fill the regions beyond $R_{2}$ with uniform density equal to
the  mean density of the galaxy distribution in the volume-limited catalog,
$n = 0.005 \ {\rm h^{3}Mpc^{-3}}$.
We smooth the galaxy density field using a Gaussian filter of radius
$R_{s} = 4 \hmpc$.
We account for boundary effects in computing the smoothed density field
$\rho_{sm}({\bf r})$ by using the ratio smoothing method of 
Melott \& Dominik (1993),
\be
\rho_{sm}({\bf r}) = \frac{\int M({\bf r'})\rho({\bf r'})W({\bf r-r'})d^{3}{\bf r'}}{\int M({\bf r'})W({\bf r - r'})d^{3}{\bf r'}} ,
\label{eqn:smmask}
\ee
where $W({\bf r})$ is the smoothing filter, and the mask array 
$M({\bf r})$ is set to 1 for pixels inside the survey region and
to 0 for pixels outside the survey region.
The rms amplitude of the  galaxy density field smoothed with a
Gaussian filter of radius $R_{s}  = 4 \hmpc$ is $\sigma_{4G} = 0.85$.
\newline
{\it Step 4: }
Monotonically map the galaxy density field onto a theoretically
determined PDF of the underlying mass distribution.
In an unbiased model, the galaxy density field is identical to the mass 
density field, so we skip this step entirely.
In a biased model, we derive the mass density field using the PDF mapping
procedure described in NW98. 
We first assume a  value for the bias factor $b$ and 
estimate the rms linear mass fluctuation using the equation
\be
\sigma_{8m} =  \frac{\sigma_{8g}}{b},
\label{eqn:bdef}
\ee
where $\sigma_{8g}$ and $\sigma_{8m}$ are the rms fluctuations in 
$8h^{-1}$Mpc spheres in the non-linear galaxy density field and the linear 
mass density field, respectively.
We use an N-body code to evolve forward in time an ensemble of initial mass 
density fields, all drawn from the same assumed power spectrum,
and all normalized to this value of $\sigma_{8m}$.
We then derive an ensemble-averaged PDF of the smoothed final mass 
density fields from the evolved mass distributions of these simulations.
While reconstructing the PSCz catalog using a model that corresponds to this
value of $\sigma_{8m}$, 
we derive a smoothed final mass density field  by  monotonically mapping 
the smoothed PSCz galaxy density field to this average PDF.
This step implicitly derives and corrects for the only monotonic 
local biasing relation that is simultaneously consistent with the observed
galaxy PDF, the assumed shape of the power spectrum and value of $b$,
 and the assumption of Gaussian initial conditions.
\newline
{\it Step 5: }
Correct for the effects of redshift-space distortions.
The peculiar velocities of galaxies distort the mapping  of galaxy positions 
from real space to redshift space, making the line-of-sight a preferred
direction in an otherwise isotropic universe.
Since we need the real space mass density field to recover the initial
mass density fluctuations, we need to correct for these redshift-space 
distortions.
On small scales, the velocity dispersion of a cluster  stretches it along
the line-of-sight into a ``Finger of God'' feature that points directly
toward the observer (e.g., \cite{delapparent86}), thereby reducing the 
amplitude of small-scale clustering.
To correct for this effect, we first identify the clusters in redshift space 
using a friends-of-friends algorithm that uses a transverse linking length 
of $0.6 \hmpc$ and a radial linking length of $500 \kms$
(\cite{huchra82}; \cite{nw87}; Moore, Frenk, \& White 1993; 
Gramann, Cen, \& Gott 1994).
For each cluster, we then shift the radial locations of its member galaxies 
so that the resulting compressed cluster has a radial velocity dispersion of 
$100 \kms$, roughly the value expected from Hubble flow across its radial
extent.
On large scales, the distortions arise from coherent  inflows into
overdense regions and outflows from underdense regions (\cite{sargent77};
\cite{kaiser87}).
To remove these distortions, we apply the following iterative procedure, which
is a modified version of the method suggested by Yahil et al. (1991) 
and Gramann et al. (1994).
This method is described in detail in NW98, and we give only a brief 
outline here.
After deriving the mass density field in step (4), we predict the velocity 
field using the second order perturbation theory relation (\cite{gr93b}),
\be
{\bf v(r)} = f(\Omega_{m})H\left[{\bf g(r)}+\frac{4}{7}\nabla C_{g}({\bf r})\right],
\label{eqn:vdelta2}
\ee
where ${\bf g(r)}$ is the gravitational acceleration field computed from 
the equation
$\nabla \cdot {\bf g(r)} = -\delta_{m}({\bf r})$ and $C_{g}$ is the solution 
of the Poisson type equation
\be
\nabla ^{2}C_{g} = \sum_{i=1}^{i=3}\sum_{j=i+1}^{j=3}\left[
  {\partial^2 \phi_g \over \partial x_i^2} 
  {\partial^2 \phi_g \over \partial x_j^2} - 
  {\left(\partial^2 \phi_g \over \partial x_i \partial x_j \right)^2} 
  \right] .
\label{eqn:cgdef}
\ee
Equation~(\ref{eqn:vdelta2}) requires that we assume a value of $\Omega_{m}$, 
to  compute the factor $f(\Omega_{m}) \approx \Omega_{m}^{0.6}$ (\cite{lss80}).
Finally, we correct the positions of galaxies so that their new positions are
consistent with their Hubble flow and the peculiar velocities at their
new locations.
We repeat these three steps until the corrections to the galaxy positions
become negligible, which usually occurs within 3 iterations.
\newline
{\it Step 6: }
Apply the dynamical \rec scheme to evolve the fluctuations backward in time.
We compute the gravitational potential from this smoothed mass
density field using the Poisson equation, then
evolve this gravitational potential backward in time using our modified
version of the Gramann (1993) method.
We use Poisson equation to derive the initial mass density fluctuations
i.e., the fluctuations that grow according to the predictions of linear theory.
\newline
{\it Step 7: }
Gaussianize this  dynamically reconstructed {\it initial}
mass density field.
This step enforces a Gaussian PDF for the initial mass density fluctuations and
yields robust reconstructions even in the non-linear regions.
\newline
{\it Step 8: }
Restore power to the recovered initial density field.
Non-linear gravitational evolution tends to suppress the small-scale power
in the reconstructed density field, beyond the suppression due to the
Gaussian smoothing alone.
We correct for this effect using the ``power restoration'' procedure
described in Weinberg (1992).
Using an ensemble of numerical simulations, we compute a set of 
correction factors $C(k)$ defined by
\be
C(k) = \left[\frac{P_{r}(k)}{P_{i}(k)}\right]^{1/2}~,
\label{eqn:ckdef}
\ee
where $P_i(k)$ is the power spectrum of a simulation's smoothed
initial conditions, and $P_r(k)$ is the power spectrum of the density
field recovered by the hybrid reconstruction method.
We multiply each Fourier mode of the reconstructed density field by $C(k)$
and also multiply by $\exp(k^2 R_s^2/2)$ in order to remove the effect of 
the original Gaussian smoothing.
Above some wavenumber $k_{\rm corr} \approx \pi/R_{\rm nl}$, where $R_{\rm nl}$
is the scale on which the rms fluctuations are unity, non-linear evolution
erases the information about the phases in the initial density field 
(\cite{lwp91}; \cite{rg91}) to the point that the hybrid method cannot 
recover it.
For $k_{\rm corr} < k \leq k_{\rm Nyq}$, therefore, we add random phase 
Fourier modes with an assumed shape for the power spectrum, where
$k_{\rm Nyq}$ is the Nyquist frequency of the grid on which we define the 
density fields.
We normalize this power spectrum by fitting it to the power spectrum of the 
recovered initial density field in the range of wavenumbers 
$k_{1} \leq k \leq k_{2} $, where $k_{1}$ and $k_{2}$ are wavenumbers
in the linear regime, with $k_{1} < k_{2} \leq k_{\rm corr}$.
In the range of wavenumbers $k_{2} < k \leq k_{\rm corr}$, multiplication
by the large factor $\exp(k^2 R_s^2/2)$ can distort the shape of the
power spectrum, although this range of wavenumbers is only in the 
mildly non-linear regime.
In this range of Fourier modes, therefore, we preserve the phases of the 
recovered initial density field but fix the amplitude of the modes to be that
determined by the fitting procedure.
In all of our simulations, $k_{\rm Nyq} = 50k_{f}$, and we choose 
$k_{\rm corr} = 15k_{f}$, $k_{1} = 4k_{f}$ and  $k_{2} = 8k_{f}$, where
$k_{f} = 2\pi/L_{\rm box} = 0.0314 \ h$Mpc$^{-1}$ is the fundamental 
frequency of the simulation box.
We assume that the shape of the power spectrum is governed by the 
parameter $\Gamma$, which is equal to $\Omega_{m}h$ in cold dark matter
models with small baryon fraction and scale-invariant inflationary
fluctuations (Efstathiou, Bond, \& White, 1992).
We do not add the small scale power using the technique of 
constrained realizations 
(\cite{bertschinger87}; \cite{hoffman91}; \cite{vandeweygaert96}) 
because we find it does not lead to a more accurate
reconstruction even for a dense sampling of the constraints 
(see NW98 for a more detailed discussion).
\newline
{\it Step 9: }
Evolve the power-restored density field forward in time
using an N-body code.
We evolve the reconstructed initial mass distribution using a particle-mesh 
(PM) code, assuming the values of $\Omega_{m}$ and $\Omega_{\Lambda}$.
This code is described and tested in  Park (1990).
We use $100^{3}$ particles and a $200^{3}$ force mesh in the PM simulations.
We start the gravitational evolution from a redshift $z = 23$ and follow
it to $z = 0$ in 46 equal incremental steps of the expansion scale factor 
$a(t)$.
We fix the amplitude of the linear mass density fluctuations 
to be $\sigma_{8m} = \sigma_{8g}/b$, where $b$ is the bias factor.
In the case of truly unbiased models (as opposed to biased models with
$b = 1.0$), we instead fix the amplitude of the linear mass fluctuations by
requiring that the non-linear rms amplitude of fluctuations in redshift
space smoothed with a Gaussian filter of radius $4 \hmpc$ ($\sigma_{4G,g}$) 
from the simulation match the observed value.
\newline
{\it Step 10: }
Compare the evolved distribution with the original galaxy distribution,
either assuming that galaxies trace mass or
using a local biasing model to select galaxies from the mass distribution.
In biased reconstructions, we choose the free parameter controlling the 
strength of the bias by requiring that the rms fluctuation 
$\sigma_{4G}$ of the reconstructed, redshift-space galaxy density field, 
smoothed with a Gaussian filter of radius $4 \hmpc$, match that of the 
original galaxy density field.

Figure 1 illustrates the intermediate steps in a hybrid \rec analysis of the 
\pscz catalog.
Panel (a) shows the redshift-space positions of all galaxies in the 
volume-limited \pscz catalog, in a slice $30 \hmpc$ thick centered on the
Supergalactic plane (SGP).
Panel (b) shows a slice through the SGP of the galaxy density field smoothed 
with a $4 \hmpc$ Gaussian filter.
The smoothed initial density field recovered by the hybrid
\rec method is shown in panel (c).
The mass distribution obtained by evolving the power-restored initial mass
density field using an N-body code is shown in panel (d).
The reconstructed, smoothed, redshift-space galaxy density field,
and the reconstructed, redshift-space galaxy distribution obtained by 
selecting galaxies from the evolved mass distribution using a power-law 
biasing model (see \S3.1 below) are shown in panels (e) and (f), respectively.
The \rec illustrated in Figure 1 assumes
$\Omega_{m} = 0.4$, $\Omega_{\Lambda} = 0.6$, and $b=0.64$.

\begin{figure}
\centerline{
\epsfxsize=5.5truein
\epsfbox[40 125 565 735]{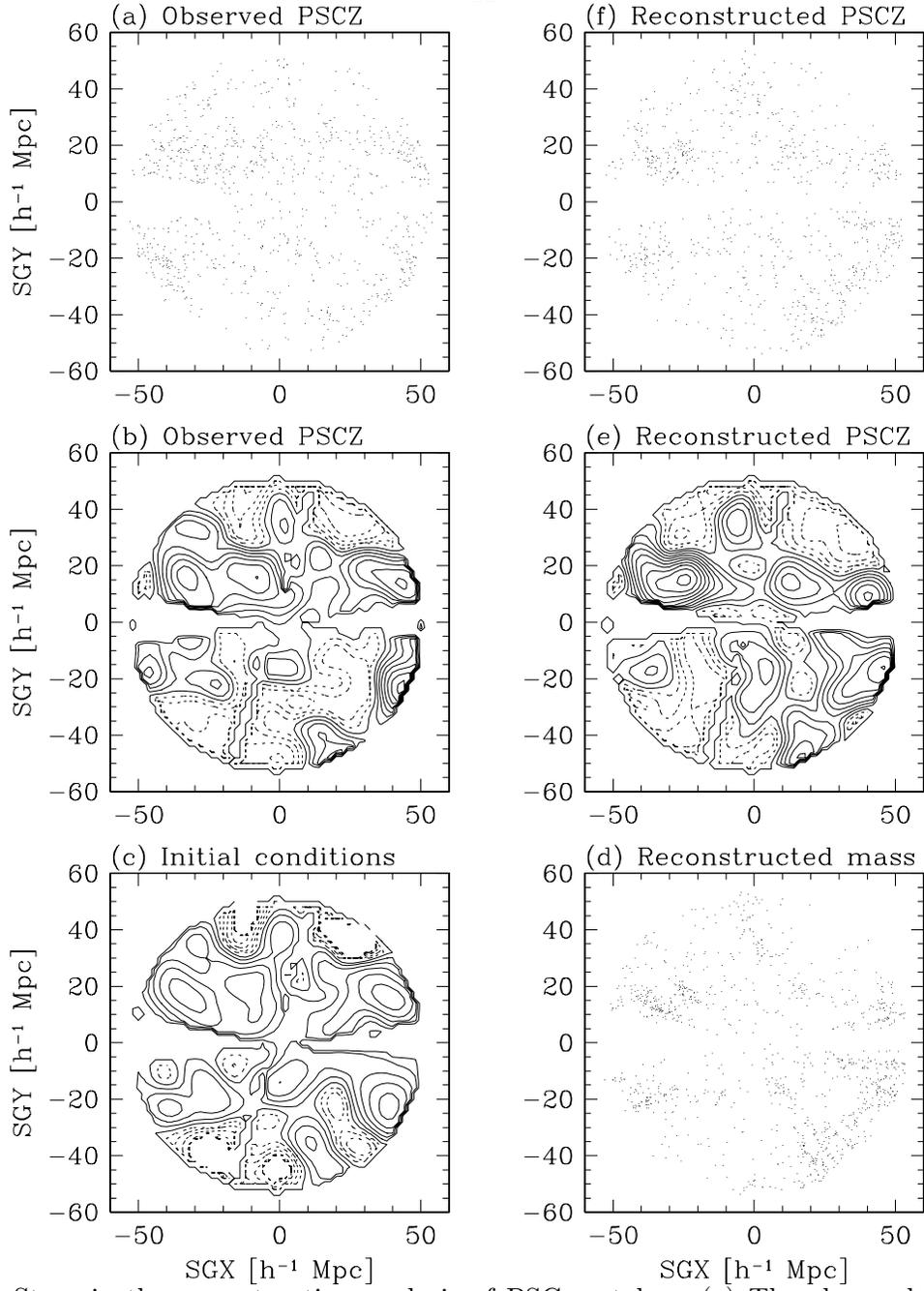}
}
\caption{Steps in the reconstruction analysis of PSCz catalog.
({\it a}) The observed PSCz galaxy distribution, 
({\it b}) the observed PSCz galaxy density field,
({\it c}) the initial mass density field linearly extrapolated to the present
epoch,
({\it d}) the reconstructed mass distribution,
({\it e}) the reconstructed PSCz galaxy density field, and
({\it f}) the reconstructed PSCz galaxy distribution.
The contour plots show the density field in the Supergalactic plane,
smoothed with a Gaussian filter of radius $R_{\rm s} = 4 \hmpc$.
Solid contours correspond to regions above the mean density $\bar \rho$
and are in steps of 0.1 in $\log(\rho/\bar \rho)$, while dashed contours 
correspond to regions below the mean density and are in steps of $0.2$ in 
$(\rho/\bar \rho)$.
The galaxy distributions show all the galaxies in redshift space in a slice 
$30 h^{-1}$Mpc thick centered on the Supergalactic plane, and volume limited 
to a radius $R_{1} = 50\hmpc$.
This reconstruction corresponds to the model L4PLb0.64 listed in Table 1.
\label{fig:scheme}
}
\end{figure}

\section{PSCz Reconstruction: Models and Mock Catalogs}

\subsection{Model Assumptions}

We use 15 different models to reconstruct the \pscz catalog
and to create mock catalogs for calibrating \rec errors.
Each model consists of a set of assumptions regarding the values of 
$\Omega_{m}$ and $\Omega_{\Lambda}$, the shape of the linear mass 
power spectrum $P(k)$ as characterized by the parameter $\Gamma$ described 
in step (8) of \S2.3, the bias factor $b$ defined as 
$b = b_{8} = \sigma_{8g}/\sigma_{8m}$, 
and the functional form of the biasing relation between IRAS galaxies and 
the underlying dark matter distribution.
As discussed in \S2.3, these assumptions influence the \rec analysis in 
different ways.
The value of $\Omega_{m}$ is required when we correct the input data 
for redshift-space distortions and when we evolve the power-restored 
initial conditions forward in time.
The shape ($\Gamma$) and amplitude ($\sigma_{8m} = \sigma_{8g}/b$) of the mass 
power spectrum are used to calculate the correction factors $C(k)$, and the 
shape is used to extrapolate the recovered initial power spectrum above 
the wavenumber $k > k_{\rm corr}$.
The value of the bias factor $b$ is required when we map the biased 
galaxy density field to the numerically determined PDF of the underlying 
mass density field with rms fluctuation amplitude 
$\sigma_{8m} = \sigma_{8g}/b$.
It is also required in the forward evolution step when we evolve
the power-restored initial mass density field to match the rms fluctuation
amplitude $\sigma_{8m}$.
We need to assume an explicit biasing scheme to select galaxies from the
evolved mass distribution.
Most of our biasing schemes have one free parameter that
we fix so that the rms fluctuation of the resulting 
galaxy distribution matches that of the input galaxy distribution, and a 
random sampling factor that we use to match the number density of galaxies
in our volume-limited \pscz sample.

Our models span a wide range of cosmological and galaxy formation parameters,
varying with respect to the following properties:
\begin{description}
\item[{(1)}: ] $\Omega_{m}$ and $\Omega_{\Lambda}$: Our assumptions for the
geometry of the background universe include Einstein de-Sitter models 
$(\Omega_{m} = 1.0,\ \Omega_{\Lambda}=0)$, open models 
$(\Omega_{m} < 1.0,\ \Omega_{\Lambda}=0)$, and flat models with 
a non-zero cosmological constant 
$(\Omega_{m} < 1.0,\ \Omega_{m} + \Omega_{\Lambda} = 1.0)$.
\item[{(2)}:] Normalization and shape of the power spectrum: 
We normalize the amplitude of the primordial mass density fluctuations 
(characterized by $\sigma_{8m}$) either to be consistent with the level 
of anisotropies in the cosmic microwave background measured by the 
COBE satellite (the COBE normalization, \cite{smoot92}) or to produce 
the observed abundance of clusters at the present epoch (the cluster 
normalization, 
White, Efstathiou \& Frenk 1993; Eke, Cole \& Frenk 1996; \cite{viana96}).
For all cluster-normalized models, we choose the values of $\Omega_{m}$
and $\sigma_{8m}$ so that $\sigma_{8m}\Omega_{m}^{0.6} = 0.55$ (\cite{wef93}), 
for both open and flat universes.
We refer the reader to Cole et al. (1997) and 
Cole et al. (1998, hereafter CHWF98) for more details of the COBE and cluster 
normalization procedures.
We define the shape parameter of the transfer function in the linear mass
power spectrum via the parameter $\Gamma$ of Efstathiou et al. (1992);
for CDM models with low baryon content, $\Gamma \approx \Omega_{m}h$.
Our power spectra include scale-invariant $(n=1)$ models with 
$\Gamma$ values consistent with the clustering properties measured from
several galaxy catalogs, viz., 
$\Gamma = 0.15 - 0.3$ (\cite{maddox90}; \cite{ebw92}; \cite{vogeley92}; 
\cite{pd94}; Gazta\~{n}aga, Croft, \& Dalton 1995; 
Maddox, Efstathiou, \& Sutherland 1996; \cite{gaztanaga98}; 
Tadros, Efstathiou, \& Dalton 1998), and some models with larger $\Gamma$ 
values.
We also consider power spectra that are normalized to both the COBE
and cluster constraints, by introducing a tilt in the spectral index of the
power spectrum.
Finally, two of our models are not normalized to COBE or clusters, although 
the rms fluctuation of the reconstructed galaxy distributions matches
that of the IRAS galaxies.
\item[{(3)}:] Bias factor: We consider models in which IRAS galaxies trace 
mass (unbiased, $b = 1.0$), models in which IRAS galaxies are more strongly 
clustered  than the mass (biased, $b > 1.0$), and models in which 
galaxies are more weakly clustered than the mass (antibiased, $b < 1.0$).
We even consider one biasing model in which the galaxies do not trace mass
but $b=1.0$, i.e., the rms amplitude of fluctuations in the galaxy and mass 
distributions are identical at the scale of $8 \hmpc$, but the galaxy density
has a non-linear dependence on the mass density.
\item[{(4)}:] Biasing scheme: Our biasing relations cover a wide range of 
plausible functional forms, with the only constraint being that they remain 
monotonic.
These include functions derived empirically from observations of different
types of galaxies, functions predicted from semi-analytic models
of galaxy formation, functions that fit the results of numerical 
studies of galaxy formation, and functions constructed ad-hoc.
All of our biasing models are deterministic, and are ``local'' in the sense 
that the efficiency of galaxy formation is determined by the properties of 
the local environment, 
i.e., by the properties within approximately one correlation length of the 
location of the galaxy.
We compute all the local properties of the mass distribution in a 
sphere of radius $4 \hmpc$ centered on the galaxy.
\end{description}

The specific biasing schemes that we use to select the IRAS galaxies from the 
evolved mass distributions of our reconstructions are as follows.
\newline
{\it Power-law bias: } 
In this simple biasing model, the IRAS galaxy density 
($\rho_{g}$) is a steadily increasing, power-law function of  the local mass 
density, 
$(\overdeng) \propto (\overdenm)^B$.  
The probability for an N-body particle with local mass density $\rho_m$ to
be selected as an IRAS galaxy is therefore
\be
P = A(\overdenm)^{B-1} .
\label{eqn:plbias}
\ee
We choose the values of $A$ and $B$ to reproduce the required number 
density and the rms fluctuation of the resulting galaxy distribution, 
respectively.
This biasing relation is similar to the one suggested by Cen \& Ostriker 
(1993) based on hydrodynamic simulations incorporating physical models
for galaxy formation (\cite{co92}), but it differs in that there
is no quadratic term that saturates the biasing relation at 
high mass densities.
\newline
{\it Threshold bias: } 
In this biasing scheme, galaxy formation is entirely 
suppressed below some threshold value of mass density, and IRAS galaxies 
form with equal efficiency per unit mass in all regions above the threshold.  
This biasing scheme was adopted in some of the early numerical investigations 
of CDM models (e.g., \cite{melott86}), and it has been used extensively 
in theoretical modeling of voids and superclusters (e.g., \cite{einasto94}).
In the density-threshold bias model, the probability that a particle
with local mass density $\rho_{m}$ is selected as an IRAS galaxy is 
\be
P = \cases{ A &if $\rho_{m} \ge B,$ \cr
            0 &if  $\rho_{m} < B$. \cr }
\label{eqn:denbias}
\ee
We choose the threshold density $B$ to match the required bias 
factor $b$, and the probability $A$ to reproduce the desired galaxy number
density.
We note that, since this model preferentially populates regions of higher 
mass density, it can only lead to a bias factor greater than unity, and hence 
cannot be used when an antibias ($b < 1.0$) is required.
\newline
{\it Morphology-density bias: }
It has been known for a long time that early-type galaxies are
preferentially found in  dense environments, while late-type galaxies
dominate in less massive groups and in the field
(\cite{hubble36}; \cite{zwicky37}; \cite{abell58}).
There have been numerous efforts to quantify this connection between 
morphology and environment (e.g., \cite{dressler80}; \cite{postman84};
\cite{lahav92}; \cite{whitmore93}),
using a variety of clustering statistics including 
angular correlation functions, redshift-space correlation functions, 
and  de-projected real-space correlation functions 
(\cite{davis76}; \cite{giovanelli86}; 
\cite{loveday95}; \cite{hermit96}; \cite{guzzo97}; \cite{willmer98}).
We model the ``bias'' arising from this morphological segregation
using the morphology-density relation proposed by Postman \& Geller (1984).
Since the IRAS-selected galaxy catalogs preferentially include dusty, 
late-type spirals (\cite{soifer84}; \cite{meiksin86}; \cite{lawrence86}; 
\cite{babul90}), we select all the spiral galaxies as IRAS galaxies.
The \md relation of Postman \& Geller (1984) assigns 
morphological types to galaxies based on the densities at the locations of all
the galaxies.
Since we evolve the power-restored density field forward in time using a 
low resolution PM code and with a finite number of mass particles,
the final densities computed over spheres centered on the galaxies 
and large enough to contain significant numbers of neighbors will be 
different from the densities at the exact locations of the galaxies.
Therefore, we recast the \md relation of Postman \& Geller (1984)
in terms of the density computed in a sphere of radius $2 \hmpc$ centered 
on the galaxy.
We assign the galaxy a spiral (Sp), S0, or elliptical (E) morphological 
type, with relative probabilities $\fsp$, $\fs0$, and $\fe$ that 
depend on the density computed within a sphere of radius $2 \hmpc$.
For $\rho<\rho_F=10\rhobar$, the morphological fractions are 
$\fsp=0.7$, $\fs0=0.2$, and $\fe=0.1$.  For 
$\rho_F<\rho<\rho_C=6 \times 10^3\rhobar$,
the fractions are
\begin{eqnarray}
\fsp & = & 0.7-0.2\alpha \nonumber \\
\fe & = & 0.1+0.1\alpha \label{eqn:mdrel} \\
\fs0  & = & 1-\fsp-\fe \nonumber \\
\alpha & = & \log_{10}(\rho/\rho_F)/\log_{10}(\rho_C/\rho_F). \nonumber
\label{eqn:md}
\end{eqnarray}
For $\rho>\rho_C$, the morphological fractions saturate at
$\fsp=0.5$, $\fs0=0.3$, and $\fe=0.2$.
The ratio of rms fluctuations in $8 \hmpc$ spheres of the elliptical and 
spiral galaxy distributions selected in this manner is 1.3,
consistent with the ratio of $\sim 1.2 - 1.7$ observed between optical 
and IRAS galaxy distributions 
(Lahav, Nemiroff, \& Piran 1990; \cite{strauss92}; \cite{saunders92}; 
\cite{pd94}; Willmer, da Costa, \& Pellegrini 1998; \cite{baker98}).
This biasing scheme has no free parameters, so the resulting 
bias factor is known {\it a priori}.
\newline
{\it Square-root Exponential bias: }
We construct a biasing scheme in which the IRAS galaxy density 
field is related to the mass density field by
\be
y = A\sqrt{x}\exp(\alpha x),
\label{eqn:sqexpbias}
\ee
where $x = \overdenm$ and $y = \overdeng$ are the mass and the IRAS galaxy 
overdensities, respectively.
We choose the values of $\alpha$ and $A$ to reproduce the required
galaxy rms fluctuation and the mean number density, respectively.
This biasing relation is a monotonically increasing function for all 
$\alpha > 0$.
We include this ad-hoc biasing scheme to test the ability of \rec analysis 
to distinguish between different biasing relations with the same bias factor.
We use this biasing scheme in a model in which galaxies do not
trace the mass, although $b=1.0$.
We note that neither the Power-law bias nor the Threshold bias can lead
to $b=1.0$, for any non-trivial values of the free parameters governing 
the strength of the bias.
\newline
{\it Semi-analytic bias: }
Except for the connection between power-law bias and the simulations of
Cen \& Ostriker (1993), the biasing models described so far are not
based on theoretical models of the galaxy formation process.
Rather, they include a variety of reasonable functional forms  that could
plausibly represent the results of some more complete theory of 
biased galaxy formation.
We now consider a biasing scheme that is motivated by a physical
theory of galaxy formation namely, the semi-analytic galaxy formation model
of Benson et al. (1999; see also Cole et al. 1994, 1999).
We parameterize this biasing scheme as  follows.
We consider the luminosities and morphologies of all the galaxies selected 
by Benson et al. (1999) from the mass distribution of the $\Lambda$CDM2 
simulation 
of Jenkins et al. (1998, the VIRGO consortium).
We select as IRAS galaxies all the galaxies whose ratio of bulge to
total mass is less than $0.4$.
The rms fluctuation of the IRAS galaxy distribution selected in this manner 
($\sigma_{8g}$), is about $10\%$ smaller than that of the underlying mass 
distribution.

The solid points in Figure 2 show the mean relation between this
``IRAS'' galaxy density field and the underlying mass density field, after 
both  fields are smoothed with a top-hat filter of radius 
$R_{\rm th} = 3 \hmpc$.
The thick solid line shows an empirical fit to this mean relation using a
smoothly varying double power-law of the form
\be
y = Ax^{\alpha} \left[C + x^{(\alpha - \beta)/\gamma} \right]^{-\gamma},
\label{eqn:sabias}
\ee
where $\alpha = 2.9$, $\beta = 0.825$, $\gamma=0.4$, $C = 0.08$, $A = 1.1$,
and $x = (1 + \delta_{m})$, $y = (1 + \delta_{g})$ are the mass and the
IRAS galaxy overdensities, respectively, smoothed with a top-hat filter of 
radius $R_{\rm th} = 3 \hmpc$.
During the \rec analysis, we use this semi-analytic biasing relation to 
select the IRAS galaxies from the evolved mass distribution.
We found that the scatter around this mean relation is dominated
by shot-noise and hence ignore it in our parameterization of the
semi-analytic bias model.
This bias model does not have any free parameters, so it results in
a known bias factor.
Note that this relation was derived for a $\Lambda$CDM model with 
$\sigma_{8m} = 0.9$, although we apply it here to an open model with 
slightly lower $\sigma_{8m}$ so that it matches the $\sigma_{8g}$ of 
IRAS galaxies.

\begin{figure}
\centerline{
\epsfxsize=\hsize
\epsfbox[18 144 592 718]{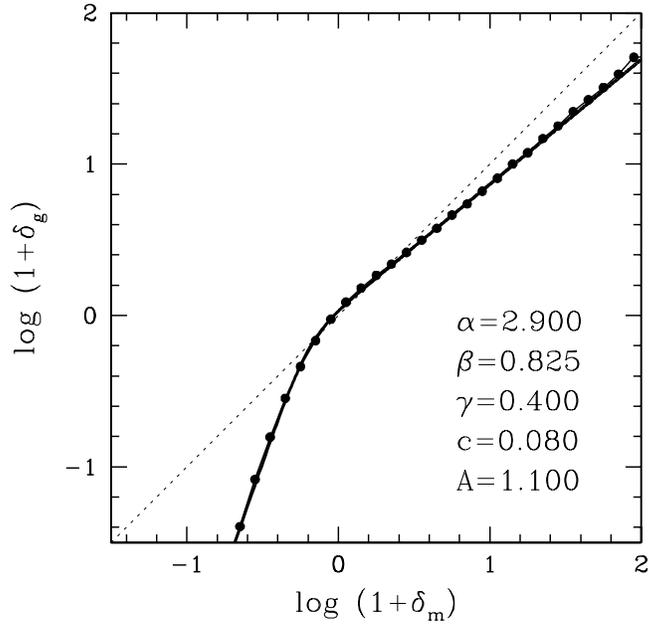}
}
\caption{The mean relation between the IRAS galaxy overdensity,
$\overdeng  = (1+\delta_{g})$, 
and the mass overdensity, $\overdenm  = (1+\delta_{m})$, in the 
Semi-analytic galaxy formation model.
Both the density fields are smoothed with a top-hat filter of radius
$R_{\rm th} = 3 \hmpc$.
The solid points show the mean relation obtained from the simulation of
Benson et al. (1999), while the thick solid line shows
the empirical fit to it using a smoothly varying double  power law,
described by equation~(\ref{eqn:sabias}).
The dotted line shows the relation $\overdeng = \overdenm$, corresponding
to an unbiased model in which galaxies trace mass.
\label{fig:sabias}
}
\end{figure}

\subsection{Models}

The 15 different models that we analyze in this paper sample the range of 
interesting values of the parameters $\Omega_{m}$ and $\sigma_{8m}$ 
(or, equivalently, the bias factor $b$).
Thus, we analyze models with $\Omega_{m} = 0.2, 0.3, 0.4, 0.5, $ and $1.0$,
while the values of $\sigma_{8m}$ range from $0.4$ to $1.44$.
The parameters of all these models are listed in Table 1.
In our model nomenclature, 
the first two symbols denote the geometry of the universe and the value of the
cosmological mass density parameter.
Thus, E1 represents an  Einstein de-Sitter model with 
$\Omega_{m} = 1.0,\ \Omega_{\Lambda}=0$;
Ox represents an open model with 
$\Omega_{m} = 0.{\rm x},\ \Omega_{\Lambda}=0$;
and Lx represents a flat model with 
$\Omega_{m} = 0.{\rm x}$ and a cosmological constant 
$\Omega_{\Lambda} =  1 - \Omega_{m}$.
The capital letters immediately following specify
the nature of the local biasing relation between the IRAS galaxy distribution
and the underlying mass distribution.
The last series of numbers after the letter $b$ corresponds to the 
bias factor of the model.
For example, the model O4SQEb1.0 specifies an open model with $\Omega_{m}=0.4$
in which the biasing relation is a square-root exponential function
and the bias factor is 1.0.

We now briefly describe the features of all the 15 models.
In \S4, we will illustrate our analysis methods and results using
6 representative models, before giving summary results for the full 
suite of 15 models in \S5.
The 6 illustrative models are:
\newline
E1UNb1.0 --- An Einstein de-Sitter universe, in which IRAS 
galaxies trace the mass distribution (unbiased), and hence $b=1.0$.
The shape of the power-spectrum is consistent with the observed clustering of 
galaxies, $\Gamma = 0.25$ (\cite{pd94}).
The mass normalization  agrees with the measured value of 
$\sigma_{8g} \approx 0.7$ for this biasing model, but it is above the
cluster normalization constraint $\sigma_{8m} = 0.55$ for $\Omega_{m} = 1$.
\newline
E1PLb1.8 --- An Einstein de-Sitter universe, in which there is a 
power-law biasing relation between the IRAS galaxy and mass 
distributions.
We choose the value of $B$ so that $b = 1.8$ and the value of $A$ so 
that $n_{g} = 0.005h^{3}$Mpc$^{-3}$.
The mass fluctuation amplitude is below the level $\sigma_{8m} = 0.55$ implied
by cluster normalization or by COBE normalization for its adopted
$\Omega_{m}$ and shape of the power spectrum.
It requires a large value of the bias factor ($b=1.8$) to match the rms
fluctuation of the IRAS galaxies.
This model has $\Gamma = 0.25$ and a tilted power spectrum with $n=0.803$;
it is similar to the E2(tilted) model of Cole et al. (1998), except that 
the rms fluctuation amplitude is $\sigma_{8m} = 0.40$ instead of 0.55.
\newline
O4MDb0.7 --- An open universe with $\Omega_{m}=0.4$ and
$\Omega_{\Lambda}=0$, in which the galaxy population as a whole traces the 
mass distribution.
We select the IRAS galaxies using the morphology-density biasing relation.
This model is cluster-normalized, and it requires the IRAS galaxies
to be antibiased with respect to the mass, i.e., $b < 1.0$.
\newline
O4SAb0.9 --- An open universe with $\Omega_{m}=0.4$ and
$\Omega_{\Lambda}=0$, in which the galaxies are selected from the mass 
distribution using the semi-analytic biasing model.
Although this model is COBE-normalized by construction, it can simultaneously 
reproduce the observed mass function of clusters (\cite{cwfr97}).
\newline
O4SQEb1.0 --- An open universe with $\Omega_{m}=0.4$ and
$\Omega_{\Lambda}=0$, in which the IRAS galaxy density field is related to the
mass density field by the square-root exponential biasing function.
We choose the value of $\alpha = 0.041$ so that $b = 1.0$ and the value of 
$A$ so that $n_{g} = 0.005h^{3}$Mpc$^{-3}$.
In this model, the IRAS galaxies {\it do not trace} the mass distribution 
even though the bias factor $b = 1.0$.
The parameters of this model are similar to those of O4SAb0.9, except that
the biasing relation is very different.
\newline
L3PLb0.62 --- A flat universe with $\Omega_{m}=0.3$ and
$\Omega_{\Lambda} = 0.7$, in which the bias between galaxies and mass
is described by the power-law bias model.
This model is COBE-normalized, and it requires the IRAS galaxies to 
be antibiased with respect to the mass distribution.
It can also reproduce the observed abundance of clusters at the present
epoch (\cite{cwfr97}).

We also reconstructed the \pscz catalog using another
set of 9 models, which, together with the 6 models described above, 
extend our exploration of the $\Omega_{m}$, $\sigma_{8m}$
parameter space.
These 9 models are all either cluster-normalized, or COBE-normalized, or both.
They are:
\newline
E1PLb1.3 --- An Einstein de-Sitter universe, in which the 
probability for a mass particle with local mass density $\rho_m$ to become 
a galaxy is given by the power-law biasing relation.
We choose the value of $B$ so that $b = 1.3$ and the value of $A$ so 
that $n_{g} = 0.005h^{3}$Mpc$^{-3}$.
This model is both cluster-normalized and COBE-normalized, and it has a tilted
power spectrum with $n = 0.803$ and $\Gamma = 0.451$.
\newline
 E1PLMDb1.3 --- An Einstein de-Sitter universe, in which the 
galaxy population as a whole is biased using a power-law function.
We choose the value of $B$ so that $\sigma_{8g} \approx 1.0$.
We then use the \md relation  to select 
all the spiral galaxies as IRAS galaxies, so that $\sigma_{8g} \approx 0.7$.
The resulting bias factor of IRAS galaxies is 
$b \approx 1.3$.
This model has the same mass power spectrum as E1PLb1.3.
\newline
 E1THb1.3 --- An Einstein de-Sitter universe, in which we select the galaxies
from the mass distribution using the threshold biasing relation.
We choose the threshold density $B$ so that the bias
factor $b = 1.3$, and the probability $A$ so that the mean galaxy density 
is $n_g=0.005\hvol$.
This model has the same mass power spectrum as E1PLb1.3.
\newline
 O2PLb0.5 --- An open universe with $\Omega_{m}=0.2$ and
$\Omega_{\Lambda}=0$, in which the IRAS galaxies are selected from the
mass distribution using the power-law biasing relation.
This model is cluster-normalized, and it has the largest amplitude of 
mass fluctuations ($\sigma_{8m} = 1.44$) among all of our 15 models.
\newline
 O3PLb1.4 --- An open universe with $\Omega_{m}=0.3$ and
$\Omega_{\Lambda}=0$, in which the IRAS galaxies are selected from the
mass distribution using the power-law biasing relation.
This is a COBE-normalized model.
\newline
 O4UNb1.0 --- An open universe with $\Omega_{m}=0.4$ and
$\Omega_{\Lambda}=0$, in which the IRAS galaxies trace the mass distribution.
Although this model is COBE-normalized by construction, it can simultaneously 
reproduce the observed mass function of galaxy clusters (\cite{cwfr97}).
\newline
 L2PLb0.77 --- A flat universe with $\Omega_{m}=0.2$ and
$\Omega_{\Lambda} = 0.8$, in which there is a power-law biasing relation
between the IRAS galaxies and the mass.
This model is COBE-normalized,  and it requires the IRAS galaxies to be
antibiased with respect to the mass.
\newline
 L4PLb0.64 --- A flat universe with $\Omega_{m}=0.4$ and
$\Omega_{\Lambda} = 0.6$, in which there is a power-law biasing relation
between the IRAS galaxies and the mass.
This model is COBE-normalized,  and it requires the IRAS galaxies to be
antibiased with respect to the mass.
This model can simultaneously reproduce the observed mass function 
of clusters (\cite{cwfr97}).
\newline
 L5PLb0.54: A flat universe with $\Omega_{m}=0.5$ and
$\Omega_{\Lambda} = 0.5$, in which there is a power-law biasing relation
 between the IRAS galaxies and the mass.
This model is also COBE-normalized, and it requires a strong antibias between
the IRAS galaxies and the mass.

\begin{table}
\caption{Parameters of the 15 models used to reconstruct the PSCz survey.
For each model, columns 1 to 7 list the name of the model, the cosmological
mass density parameter $\Omega_{m}$, the cosmological constant 
$\Omega_{\Lambda}$, the rms amplitude of the mass density fluctuations in
$8 \hmpc$ spheres $\sigma_{8m}$, the cluster normalization parameter
$\sigma_{8m}\Omega_{m}^{0.6}$, 
the parameter $\beta_{\rm IRAS} = \Omega_{m}^{0.6}/b$ where 
$b = \sigma_{8\rm IRAS}/\sigma_{8m}$ is the bias parameter of IRAS galaxies, 
and the shape parameter $\Gamma$ that specifies the shape of the linear
mass transfer function (Efstathiou et al. 1992).
A Y (N) in column 8 shows that the model satisfies (does not satisfy) the 
cluster constraint, while a similar notation in column 9 shows whether the
model is COBE normalized.
Column 10 lists the type of local biasing model used to select the IRAS 
galaxies from the underlying mass distribution.
The final column specifies the name of the simulation in CHWF98 from which we 
created the mock catalogs for each model.
}
\bigskip
\begin{tabular}{*{12}{c|}}
\tableline\tableline
& & & & & & & \multicolumn{2}{c|}{Normalization} & & Mock \\ \cline{8-9} 
Name & $\Omega_{m}$ & $\Omega_{\Lambda}$ & $\sigma_{\rm 8m} $ & $\sigma_{\rm 8m}\Omega_{m}^{0.6} $ & $\beta_{\rm IRAS}$ & $\Gamma$ & Cluster & COBE & Bias/Antibias & catalog \\
\tableline
E1UNb1.0    & 1.0 & 0.0 & 0.70 & 0.70 & 1.00 & 0.250 & N & N & Unbiased & E3S \\
E1PLb1.3    & 1.0 & 0.0 & 0.55 & 0.55 & 0.78 & 0.451\tablenotemark{\dagger} & Y & Y & Power-law & E2\\
E1PLMDb1.3  & 1.0 & 0.0 & 0.55 & 0.55 & 0.78 & 0.451\tablenotemark{\dagger} & Y & Y & Power-law/MD& E2 \\ 
E1THb1.3    & 1.0 & 0.0 & 0.55 & 0.55 & 0.78 & 0.451\tablenotemark{\dagger} & Y & Y & Threshold & E2 \\ 
E1PLb1.8    & 1.0 & 0.0 & 0.40 & 0.40 & 0.57 & 0.451\tablenotemark{\dagger} & N & N & Power-law & E2 \tablenotemark{*}\\ 
O2PLb0.5    & 0.2 & 0.0 & 1.44 & 0.55 & 0.78 & 0.25  & Y & N & Power-law & O2S \\
O3PLb1.4    & 0.3 & 0.0 & 0.5  & 0.24 & 0.34 & 0.172 & N & Y & Power-law & O3 \\
O4UNb1.0    & 0.4 & 0.0 & 0.75 & 0.43 & 0.61 & 0.234 & N & Y & Unbiased & O4 \\
O4MDb0.7    & 0.4 & 0.0 & 0.95 & 0.55 & 0.78 & 0.25  & Y & N & Morph-den & O4S \\ 
O4SAb0.9    & 0.4 & 0.0 & 0.75 & 0.43 & 0.61 & 0.234 & N & Y & Semi-analytic & O4 \\
O4SQEb1.0   & 0.4 & 0.0 & 0.75 & 0.43 & 0.61 & 0.234 & N & Y & Sq-Exp. & O4 \\
L2PLb0.77   & 0.2 & 0.8 & 0.9  & 0.34 & 0.49 & 0.131 & N & Y & Power-law & L2 \\
L3PLb0.62   & 0.3 & 0.7 & 1.13 & 0.55 & 0.78 & 0.25  & Y & N & Power-law & L3S \\
L4PLb0.64   & 0.4 & 0.6 & 1.1  & 0.63 & 0.90 & 0.213 & N & Y & Power-law & L4 \\
L5PLb0.54   & 0.5 & 0.5 & 1.3  & 0.86 & 1.23 & 0.27  & N & Y & Power-law & L5 \\
\tableline
\end{tabular}
\tablenotetext{*}{Mock catalogs for all models were drawn from the simulations
of CHWF98, except for the model E1PLb1.8, for which we created the mass 
distribution using a lower resolution PM simulation. See text for more 
details.}
\tablenotetext{\dagger}{These models have a tilted  inflationary power spectrum with
$n=0.803$, while all other models have a scale invariant ($n=1$) inflationary
power spectrum.}
\label{table:models}
\end{table}

\subsection{Mock Catalogs}

If the \rec method were perfect, and structure in the universe really did form
from gravitational instability of Gaussian initial conditions, then
we would expect to reproduce exactly the galaxy distribution in the \pscz 
catalog, if we assumed the correct value of $\Omega_{m}$ and the correct 
biasing relation between IRAS galaxies and mass.
However, the \rec method suffers from inaccuracies arising at various 
intermediate steps --- inaccuracies in the bias mapping procedure, 
inaccuracies in correcting for the redshift-space distortions, inaccuracies 
in the dynamical recovery of the initial mass density fluctuations, and 
inaccuracies in the forward evolution step caused by poor modeling of the
large scale tidal field and (on small scales) the finite numerical resolution.
All these errors accumulate at various levels, with the result that we cannot
expect a \rec to produce an exact match to the input data even if all of its
assumptions are correct.
It is therefore necessary to calibrate the magnitude of the errors intrinsic
to the \rec method before we can derive any conclusions regarding the validity
of the various assumptions entering the \rec procedure.

We assess these errors by reconstructing a set of mock 
\pscz catalogs for each of the 15 different models.
For every model, we construct the mock \pscz catalogs from the outputs of 
numerical simulations that have the appropriate values of $\Omega_{m}$ and
bias.
The geometry, the sky-coverage, the depth, and the selection function of the
mock catalogs all mimic those of the original \pscz catalog.

We construct the mock catalogs for 14 of the 15 models using the outputs of
the N-body simulations of cold dark matter models performed by CHWF98.
The CHWF98 simulations use a modified version of the AP3M code of 
Couchman (1991) to follow the gravitational evolution of $192^{3}$ particles 
in a periodic cubical box of side $345.6 \hmpc$, using a gravitational 
softening length of $\epsilon = 90 \hkpc$ (for a Plummer force law), fixed 
in comoving coordinates.
Further details of the simulations are in CHWF98.
For the model E1PLb1.8, we created the mass distribution by evolving an 
initial density field with parameters similar to the E2(tilted) model of 
CHWF98, except that $\sigma_{8m} = 0.4$ instead of 0.55.
We evolved $192^{3}$ particles on a $384^{3}$ force mesh using the PM code 
of Park (1990).
Our goal here was to investigate an $\Omega_{m} = 1.0$ model with lower 
mass fluctuation amplitude than those considered by CHWF98, which is why
we needed to run a new simulation.
For the other 14 models, the $90 \hkpc$ force resolution of the mock catalog
simulation is much higher than the $\sim 1 \hmpc$ force resolution
of the PM simulation used in the forward evolution step of the \rec procedure.
Our calibration of systematic errors therefore includes the error caused
by limited force resolution of the PM simulations.

For every model (except E1PLb1.8), the last column in Table 1 
lists the CHWF98 simulation from which we derive the mock catalogs.
If the model involves bias, we start by selecting the galaxies from
the mass distribution using the appropriate biasing algorithm.
We then select ``observers'' from the galaxy distributions so that they 
satisfy the following observed properties of the Local Group:
\begin{description}
\item[{(1)}:] The velocity of the Local Group observer should be in the
range $550 \kms < V_{LG} < 700 \kms$, consistent with the  amplitude
of the dipole anisotropy in the cosmic microwave background (\cite{smoot91}).
\item[{(2)}:] The overdensity of galaxies in a spherical 
region of radius $5 \hmpc$ centered on the Local Group observer should
be in the range $1.0 < 1+\delta_{g}(5 \hmpc) < 2.0$ 
(\cite{brown87}; \cite{hudson93}; \cite{schlegel94}).
\item[{(3)}:] The radial velocity dispersion in a sphere of radius 
$5 \hmpc$ around the Local Group observer should be less than $150 \kms$, 
consistent with the observations of a cold velocity field near the Local Group
(\cite{sandage75}; \cite{sandage86}; \cite{giraud86}; \cite{schlegel94}).
We note that for all but one of the galaxy distributions (corresponding to 
the E1UNb1.0 model), our Local Groups have local velocity dispersion
smaller than $100 \kms$.
\item[{(4)}:] The  Local Group particles for any pair of mock catalogs 
constructed from a simulation should be separated by at least $50 \hmpc$.
This criterion ensures that the density fields in the mock \pscz catalogs 
centered on these observers are quite different from each other, at least 
within the volume-limiting radius $R_{1} = 50 \hmpc$.
\end{description}

We assign each particle in the galaxy distribution a redshift based on its 
real space distance and its radial peculiar velocity with respect to 
the Local Group observer particle.
We assign luminosities to these galaxies consistent with the luminosity
function of the IRAS galaxies in the \pscz catalog.
We ``observe'' this galaxy distribution using the selection function of the 
\pscz survey.
We reject all the galaxies in the angular regions not covered by 
the \pscz catalog, so that the sky coverage in the mock catalogs is identical 
to that of the true \pscz  catalog.
We create 10 mock \pscz catalogs for each of the 15 models and 
reconstruct them in exactly the same manner as the \pscz catalog.

\section{PSCz Reconstruction: Illustrative Results}

We now describe the results of reconstructing the \pscz catalog and ten mock
catalogs for the first 6 of the 15 models described in \S3.2.
Figure 3 shows a slice through the galaxy density fields of the true and the
reconstructed \pscz catalogs.
The density fields have been convolved with a Gaussian filter 
$e^{-r^{2}/2R_{s}^{2}}$, with smoothing radius $R_{s}=4h^{-1}$Mpc.
The slices show the contours of the density field in the SGP.
The galaxy density field traced by the galaxies in the \pscz catalog
is shown in panel (a).
Some of the prominent features include the Perseus-Pisces supercluster
seen as the overdensity near the boundaries in the bottom right region,
the Great Attractor region in the diagonally opposite direction near the top 
left corner, and the Local void in the bottom left region.
We refer the reader to Branchini et al. (1999) for a detailed cosmographical
description of the  \pscz catalog.
Panels (b) through (f) show the galaxy density fields reconstructed in the
models E1UNb1.0, E1PLb1.8, O4MDb0.7, O4SAb0.9, and L3PLb0.62, respectively.
All the models can, at least qualitatively, 
reproduce  the general features of the observed \pscz galaxy distribution.
This success offers support to the hypothesis that structure formed from the
gravitational instability of Gaussian primordial mass density fluctuations.
We will see below that, although the various reconstructions resemble 
the observed \pscz galaxy density field in this visual representation, there 
are quantifiable differences between the accuracy of the reconstructions 
corresponding to different models.
Thus, some models (like, for example, the models O4MDb0.7, O4SAb0.9 and
L3PLb0.62) can reconstruct the \pscz catalog as well as can be 
expected based on the mock catalog reconstructions, while others (including the
models E1UNb1.0 and  E1PLb1.8) fail in a systematic manner.

\begin{figure}
\centerline{
\epsfxsize=5.5truein
\epsfbox[40 50 565 735]{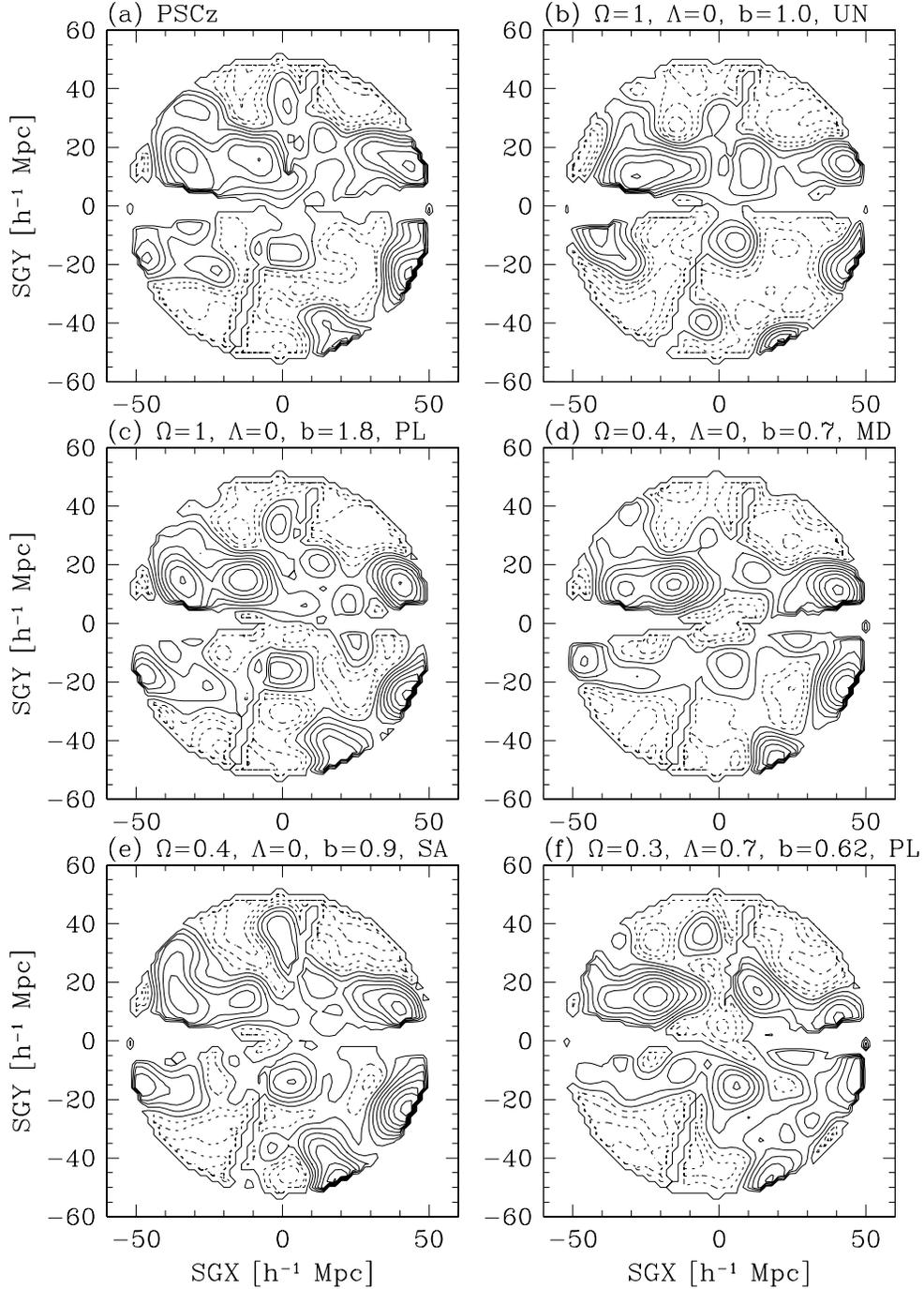}
}
\caption{Contours in the Supergalactic plane in the true and reconstructed 
PSCz density fields.
The density fields are smoothed with a Gaussian filter of radius 
$R_{s} = 4h^{-1}$Mpc.
Solid contours correspond to regions above the mean density $\bar \rho$
and are in steps of 0.1 in $\log(\rho/\bar \rho)$, while dashed contours 
correspond to regions below the mean density and are in steps of $0.2$ in 
$(\rho/\bar \rho)$.
({\it a}) The observed PSCz density field.
Remaining panels show the density field reconstructed in the following models:
({\it b}) E1UNb1.0, ({\it c}) E1PLb1.8, ({\it d}) O4MDb0.7,  
({\it e}) O4SAb0.9,  and ({\it f}) L3PLb0.62.
\label{fig:slicef}
}
\end{figure}

Figure 4 shows the redshift-space locations of galaxies in the 
volume-limited \pscz catalog and its reconstructions.
We plot the SGX and SGY coordinates of all the galaxies that lie
in a slice $30 h^{-1}$Mpc thick centered on the SGP.
The different panels show the true \pscz galaxy distribution and the
various reconstructions, in the same format as Figure 3.

\begin{figure}
\centerline{
\epsfxsize=5.5truein
\epsfbox[40 50 565 735]{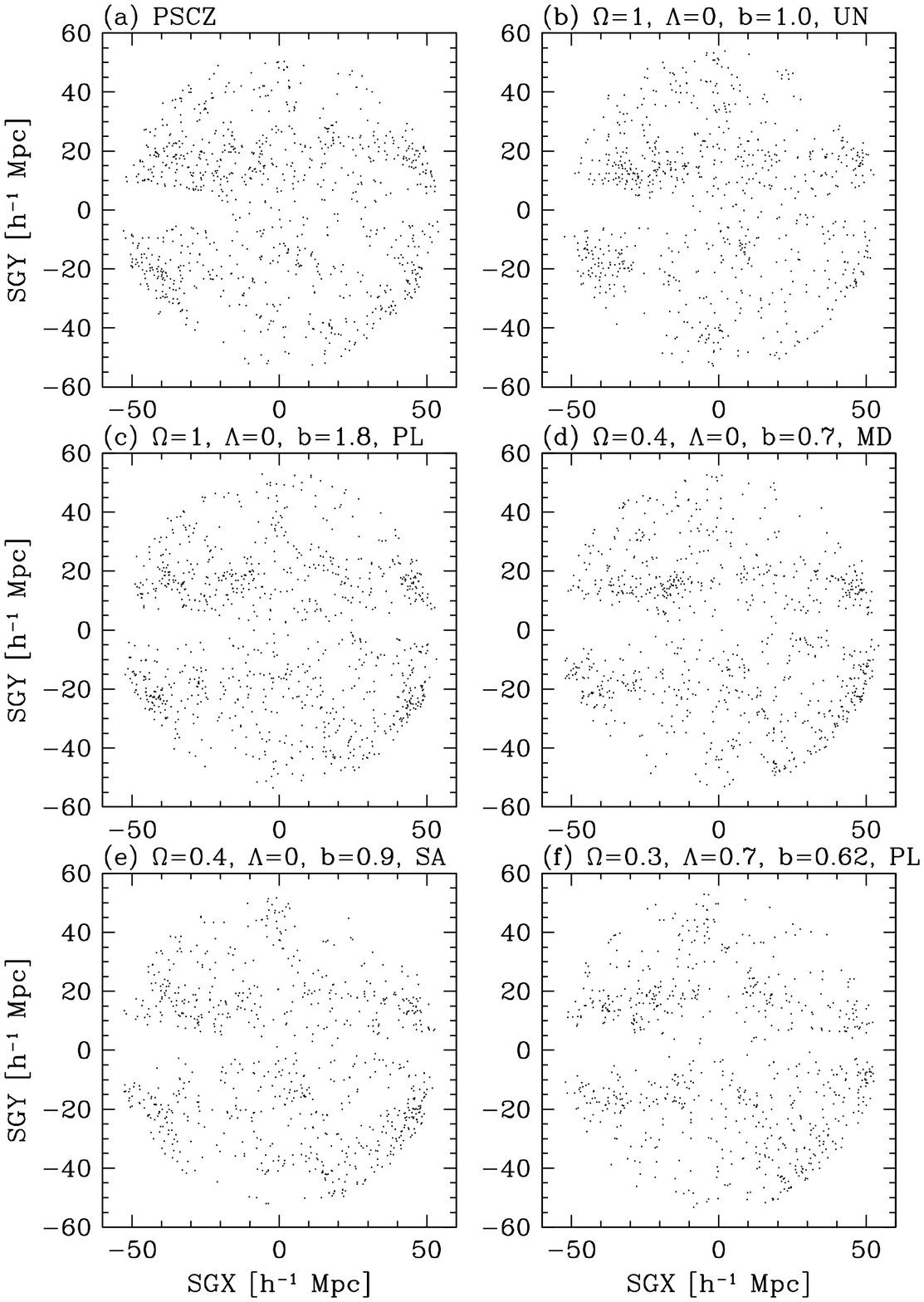}
}
\caption{Galaxy distributions in redshift space in the PSCz survey and its
reconstructions.
The panels show all the galaxies in a slice $30 h^{-1}$Mpc thick centered on 
the Supergalactic plane, and volume limited to a radius $R_{1} = 50\hmpc$.
({\it a}) The observed PSCz galaxy distribution.
Remaining panels show the reconstructed galaxy distributions for the models 
listed in Fig.\ 1.
\label{fig:zpart}
}
\end{figure}

One of the most obvious quantitative measurements of the success of a \rec
is the correlation coefficient $r$ between the original and the
reconstructed smoothed galaxy density fields,
\be
r \equiv \frac{\left< \delta_{r}\delta_{t}\right>}{\left<\delta_{r}^{2}\right>^{\frac{1}{2}}\left<\delta_{t}^{2}\right>^{\frac{1}{2}}},
\label{eqn:rdef}
\ee
where $\delta_{t}$ and $\delta_{r}$  are respectively the original and 
reconstructed  smoothed galaxy density fields.
Panels (a) through (f) of Figure 5 show the correlation coefficients for the 
models E1UNb1.0, E1PLb1.8, O4MDb0.7, O4SAb0.9, O4SQEb1.0, and L3PLb0.62, 
respectively.
For every model, we assign ranks to the reconstructions of each of the 10 
mock catalogs and to the \rec of the true \pscz catalog, in descending order 
of their values of $r$: 
the catalog whose \rec has the highest $r$ value is assigned a rank of 0,
the catalog whose \rec has the lowest $r$ value is assigned a rank of 10,
and so on in between.
The solid line in each panel shows the values of $r$ for the 10 mock
catalog reconstructions of the model, in rank order.
The horizontal dashed line shows the value of $r$ for the \pscz \rec based on
the model assumptions.
We find that the absolute values of $r$ tend to decrease for models with
larger values of $\sigma_{8m}$ (smaller values of $b$) 
because the greater degree of non-linear gravitational evolution makes
the recovery of initial conditions less accurate.
The {\it absolute} value of $r$ is therefore of little use for comparing
the viability of different reconstruction models.
We focus instead on the value of $r$ relative  to the values expected 
given the model assumptions, and since the \pscz reconstructions for all 
six of these models have a rank of seven or less (and the models therefore 
reproduce \pscz better than they reproduce at least three of their own mock 
catalogs), we conclude that all of them are acceptable by this particular 
measure.

\begin{figure}
\centerline{
\epsfxsize=\hsize
\epsfbox[18 144 592 718]{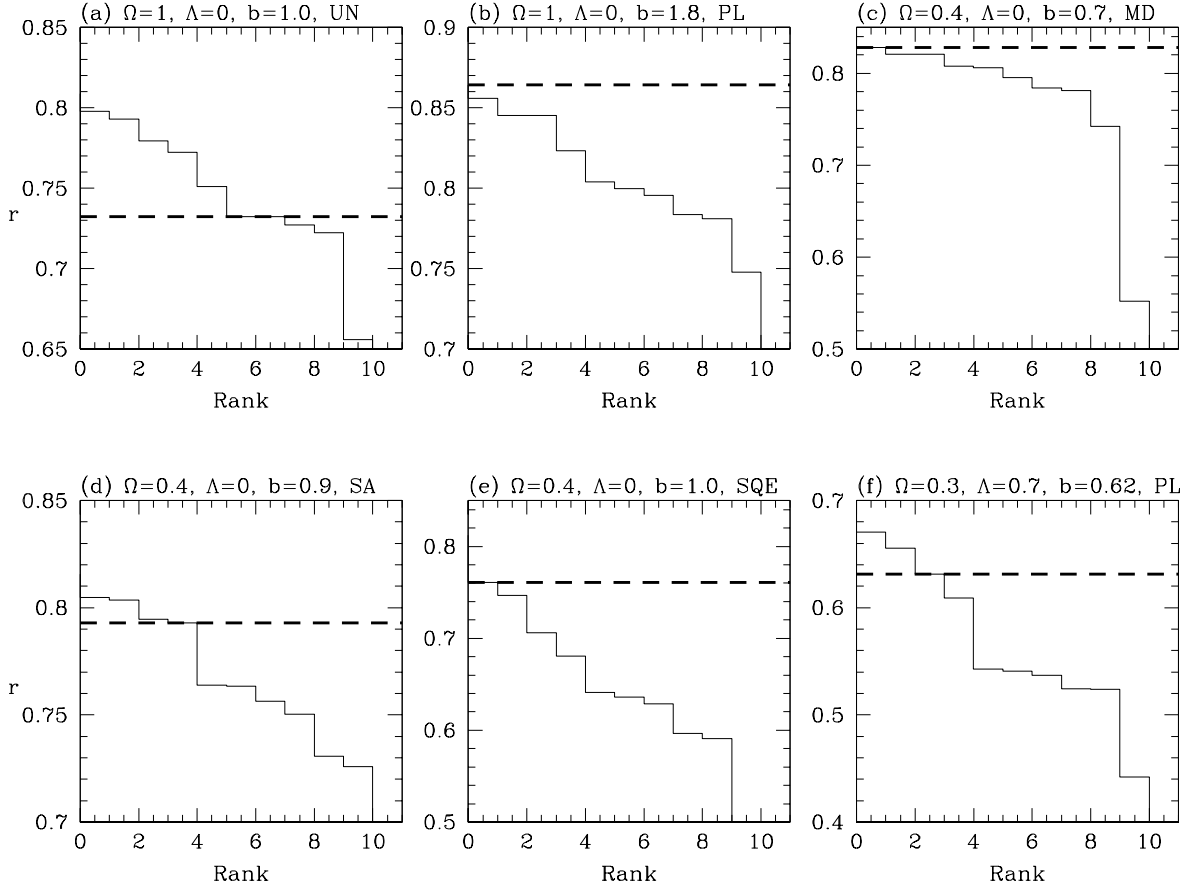}
}
\caption{The correlation between the true and reconstructed smoothed
galaxy density fields.
The solid line in all the panels shows the distribution of the correlation
coefficient (eq.~\ref{eqn:rdef}) between the true and reconstructed 
smoothed galaxy density fields
for the ten independent mock catalogs, while the 
dashed line shows the same quantity for the PSCz reconstruction.
The different panels show the galaxy distributions reconstructed with the 
following assumptions:
({\it a}) E1UNb1.0, ({\it b}) E1PLb1.8, ({\it c}) O4MDb0.7,   
({\it d}) O4SAb0.9, ({\it e}) O4SQEb1.0, and ({\it f}) L3PLb0.62.
Note that the range of the vertical axis varies from one panel to another.
\label{fig:rxy}
}
\end{figure}

The correlation coefficient quantifies the success of the \rec in 
matching the observed galaxy density field at a particular scale,
$R_{s} = 4 \hmpc$.
In order to probe a range of scales, Figure 6 shows the distribution of the 
Fourier difference statistic
\be
D(k) \equiv \frac{\sum \vert \tilde \delta_{r}({\bf k}) - \tilde \delta_{t}({\bf k}) 
\vert ^{2}}{\sum \left( \vert \tilde \delta_{r}({\bf k}) \vert ^{2} + \vert \tilde \delta_{t}({\bf k}) \vert ^{2}\right)},
\label{eqn:dkdef}
\ee
where the subscripts $t$ and $r$ refer to the true and reconstructed
density fields respectively, and $\tilde \delta({\bf k})$ represents the
complex Fourier component of the density field.
The summation is over all the waves with $\vert {\bf k} \vert$ in the interval 
$\left(k-1,k\right]$.
This statistic measures the difference in both the moduli and the 
phases of the Fourier components of the true and reconstructed 
density fields, and it is independent of any smoothing of the density fields.
It was first used by Little et al. (1991) to demonstrate the 
effects of power-transfer from large scales to small scales during non-linear 
gravitational evolution.
When the complex amplitudes of the Fourier components of the 
true and reconstructed density fields are identical, $D(k) = 0$, 
while for two fields with uncorrelated phases, the average value of 
$D(k)=1$.
Figure 6 shows the value of $D(k)$ averaged over the range of
wavenumbers $k_{\rm surv} < k < k_{8}$, where 
$k_{\rm surv} = 2\pi/(2 \times R_{1}) = 0.0628 \ h$Mpc$^{-1}$ is the wavenumber
corresponding to the size of the reconstruction volume, and 
$k_{8} = 2\pi/(2 \times 8) = 0.3927 \ h$Mpc$^{-1}$ is the wavenumber 
corresponding to the length scale of $8 \hmpc$, approximately equal to 
the scale of non-linearity in the \pscz galaxy distribution.
The different panels correspond to the six models in the same format as
Figure 5.
The solid line in each panel shows the distribution of $D(k)$ in the ten mock
catalog reconstructions of that model.
The dashed line shows the value of $D(k)$ in the \pscz reconstruction.
We rank the mock catalog reconstructions and the \pscz reconstruction,
in ascending order of their values of $D(k)$.
We find that the \pscz \rec has a high rank (of 9) in the model E1UNb1.0,
and has smaller ranks in all the other models.
Hence, the model E1UNb1.0 fails (though only at the $80\%$ confidence 
level) to reproduce the \pscz density field as measured by this
statistic, while the other five models yield acceptable reconstructions.

\begin{figure}
\centerline{
\epsfxsize=\hsize
\epsfbox[18 144 592 718]{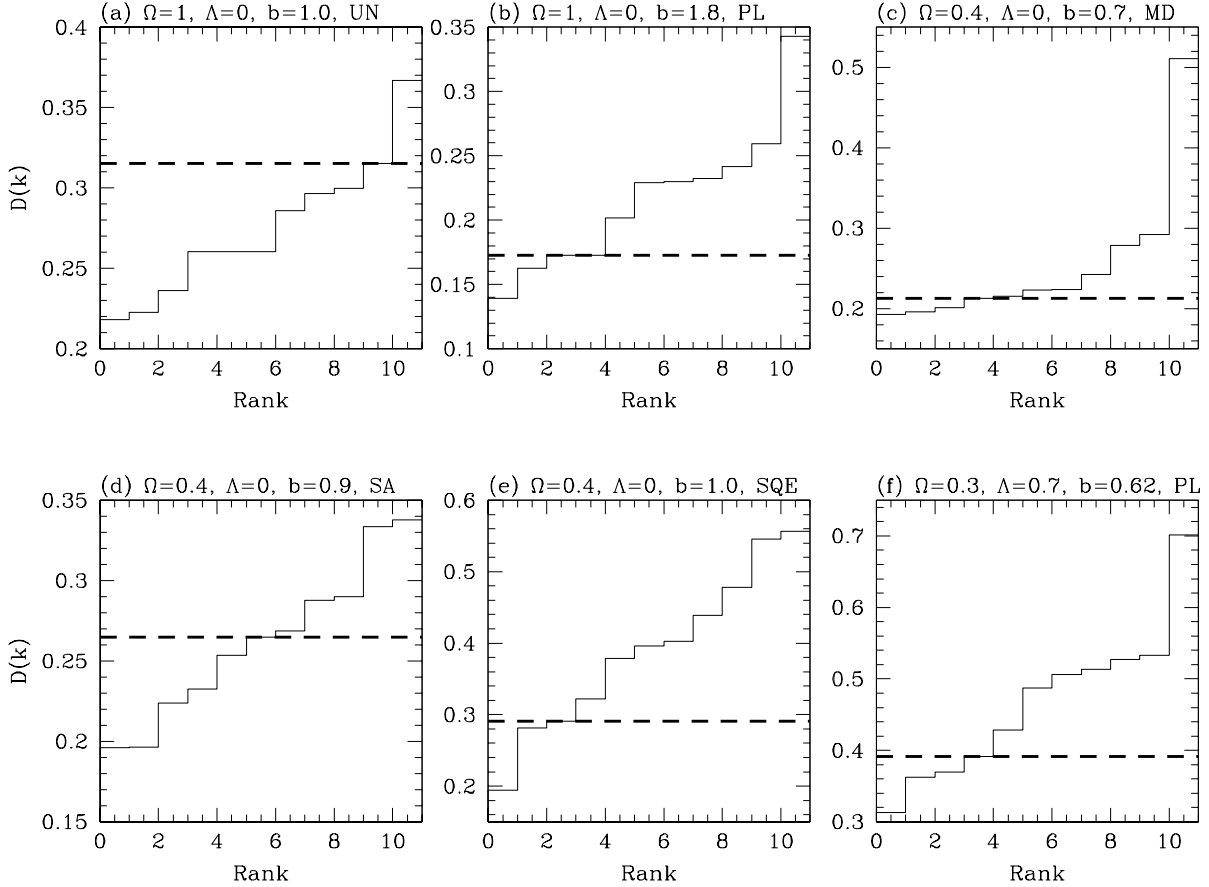}
}
\caption{The Fourier difference statistic $D(k)$ between the true and the 
reconstructed galaxy density fields (eq.~\ref{eqn:dkdef}).
The solid line in all the panels shows the distribution of the value 
of $D(k)$ averaged over the range of wavenumbers $k_{\rm surv} < k < k_{8}$ 
in the ten independent mock catalogs, while the dashed line 
shows the same quantity for the PSCz reconstruction.
The different panels correspond to the galaxy distributions reconstructed 
using the various models listed in Fig.\ 4.
Note that the axis range varies from one panel to another.
\label{fig:fdiff}
}
\end{figure}

Figure 7 shows the PDF of the true and reconstructed galaxy density fields
for the six models.
We compute the PDF of a density field after smoothing it with a Gaussian filter
of radius $R_{s} = 4 \hmpc$.
In every panel, the crosses and the thin solid line show the PDFs of the 
true and reconstructed galaxy density fields for a typical
mock catalog --- the one with rank $5$ according to the Figure-of-Merit
(FOM) for this statistic, defined as the maximum value of the absolute 
difference between the cumulative distributions $C_{t}(\nu)$ and $C_{r}(\nu)$ 
of the true and reconstructed galaxy density fields,
\be
{\rm FOM}_{\rm PDF} = {\rm max}\ \vert C_{t}(\nu) - C_{r}(\nu) \vert.
\label{eqn:fompdf}
\ee
This is the FOM that would be used in a Kolmogorov-Smirnov
comparison of the PDFs, and we find that it gives similar results
in terms of ranks to a FOM based on absolute differences of the
differential PDFs.
The open circles and the thick solid line show the true and
reconstructed PDFs for the \pscz reconstruction, offset  vertically by 
0.2 for the sake of clarity.

For every model, we rank the mock catalog reconstructions and the \pscz 
reconstruction, in increasing order of the FOM of the statistic.
If the \rec of \pscz based on the model assumptions is worse than expected
from the mock catalog tests, then the \pscz \rec will have a high rank.
A low \pscz rank, conversely, implies a \rec that is successful given the
expectations from the mock catalog tests.
Visual comparison between the PDF recoveries for \pscz and for the
rank-$5$ mock catalogs in Figure 7 suggests that the \pscz reconstructions
for the models E1UNb1.0, E1PLb1.8, O4MDb0.7, and O4SQEb1.0 should have high 
ranks, while the \pscz reconstructions for the models O4SAb0.9 and L3PLb0.62 
should have low ranks.
This is indeed the case, as can be verified by the \pscz ranks listed in 
each panel.

We show the ranks for all other statistics in Figures 8 through 12,
in the same format as in Figure 7 for the PDF statistic.
If the \pscz catalog has a rank of 5 for any of the statistics, we 
show the results for the mock catalog ranked 6 according to that statistic.
We will use the ranks for all the statistics as the basis for a more 
systematic evaluation of models  in \S5.

\begin{figure}
\centerline{
\epsfxsize=\hsize
\epsfbox[18 200 592 718]{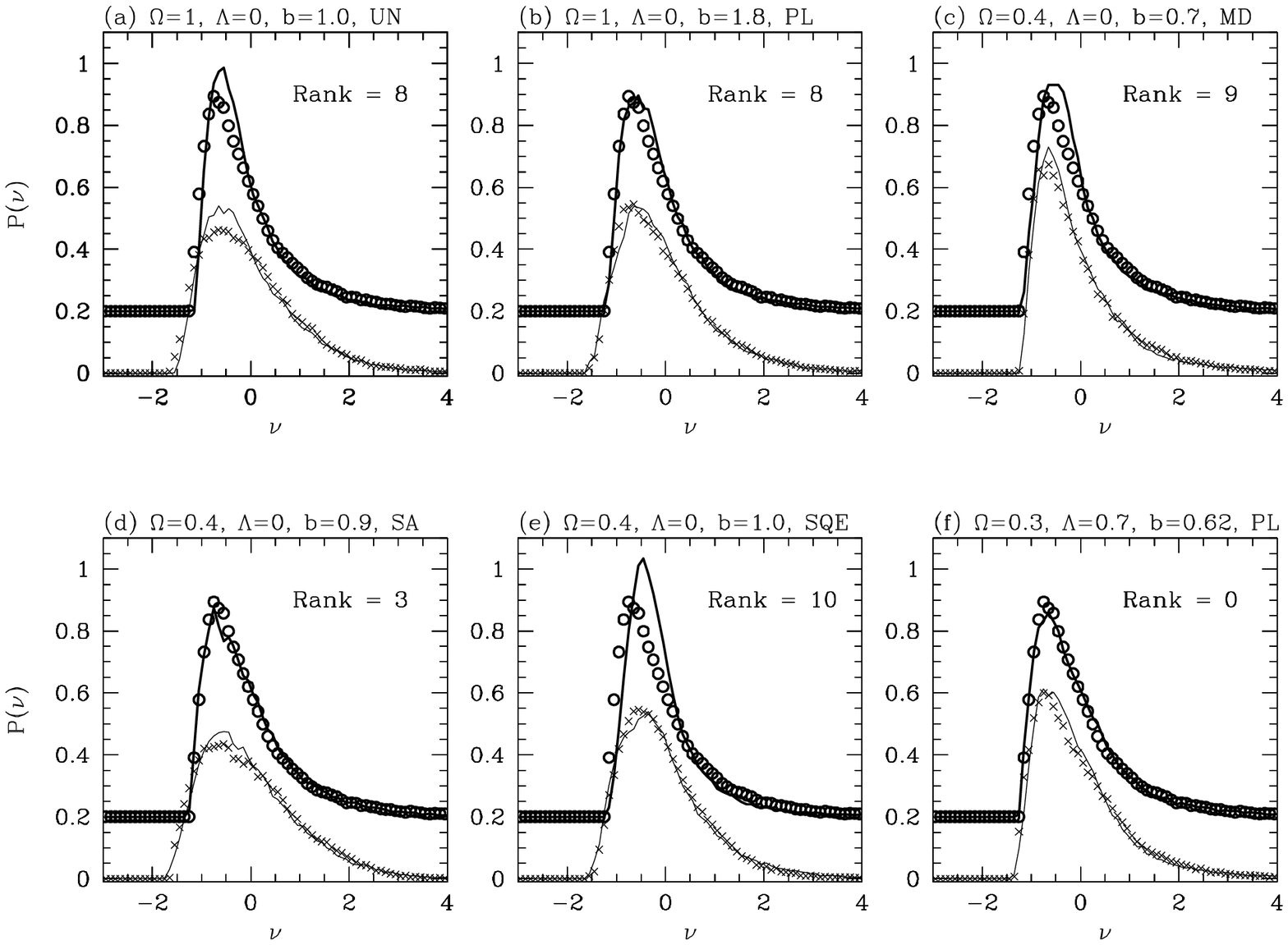}
}
\caption{PDF of the true and reconstructed galaxy density fields.
All the density fields are smoothed with a Gaussian filter of radius 
$R_{s} = 4h^{-1}$Mpc.
The crosses and the thin solid line show the PDFs of the true and 
reconstructed smoothed density fields of a mock catalog, the one with
rank 5 according to the FOM for this statistic defined in 
equation~(\ref{eqn:fompdf}).
The open circles and the thick solid line show the same quantities for the 
PSCz catalog, and are vertically offset by 0.2 for clarity.
The different panels correspond to the galaxy distributions reconstructed 
using the various models listed in Fig.\ 4.
A successful model should reproduce the PSCz (i.e., show agreement between 
the circles and thick line) about as well as it reproduces its own mock
catalogs (as illustrated for a typical case by the crosses and thin line). 
Each panel lists the rank of the \pscz \rec based on the model assumptions,
relative to the reconstructions of the model mock catalogs, according to the
FOM for this statistic.
\label{fig:pdf}
}
\end{figure}

Figure 8 shows the distribution of galaxy counts in spheres of radius 
$8 \hmpc$, in the true and reconstructed galaxy distributions for the 
six models.
We compute this distribution by placing $50,000$ spherical cells at random
locations within the reconstruction volume and counting the number of galaxies
within each cell.
If a cell lies close to the boundary of the survey region, we include it 
in the distribution only if at least $90\%$ of its volume lies within the
survey region.
The crosses and the thin solid line show the distributions of counts in
the true and reconstructed galaxy distributions of the mock catalog with
rank 5.
The open circles and the thick solid line show the same 
quantities for the PSCz catalog, and are vertically offset by 0.05 for clarity.
We define the FOM for this statistic as
\be
{\rm FOM}_{\rm COUNTS}  = \sum_{N=1}^{N=\infty} \vert P_{t}(N) - P_{r}(N) \vert,
\label{eqn:fomcounts}
\ee
where $P_{t}(N)$ and $P_{r}(N)$ are the distributions of the counts in cells
in the true and reconstructed galaxy distributions.
From visual inspection of Figure 8, we would expect the models 
E1UNb1.0, E1PLb1.8, and O4SQEb1.0 to have high ranks and the models O4MDb0.7, 
O4SAb0.9, and L3PLb0.62 to have low ranks, as is indeed verified by the ranks 
of the \pscz \rec listed in different panels.
We also computed this distribution using spherical cells of radius $3 \hmpc$,
but do not show the corresponding figure.
Although the distribution of galaxy counts is a measure similar to the PDF 
statistic shown in Figure 7, here we are using different smoothing 
filters (top-hat instead of Gaussian) and smoothing lengths
($3 \hmpc$ and $8 \hmpc$ instead of $4 \hmpc$).

\begin{figure}
\centerline{
\epsfxsize=\hsize
\epsfbox[18 144 592 718]{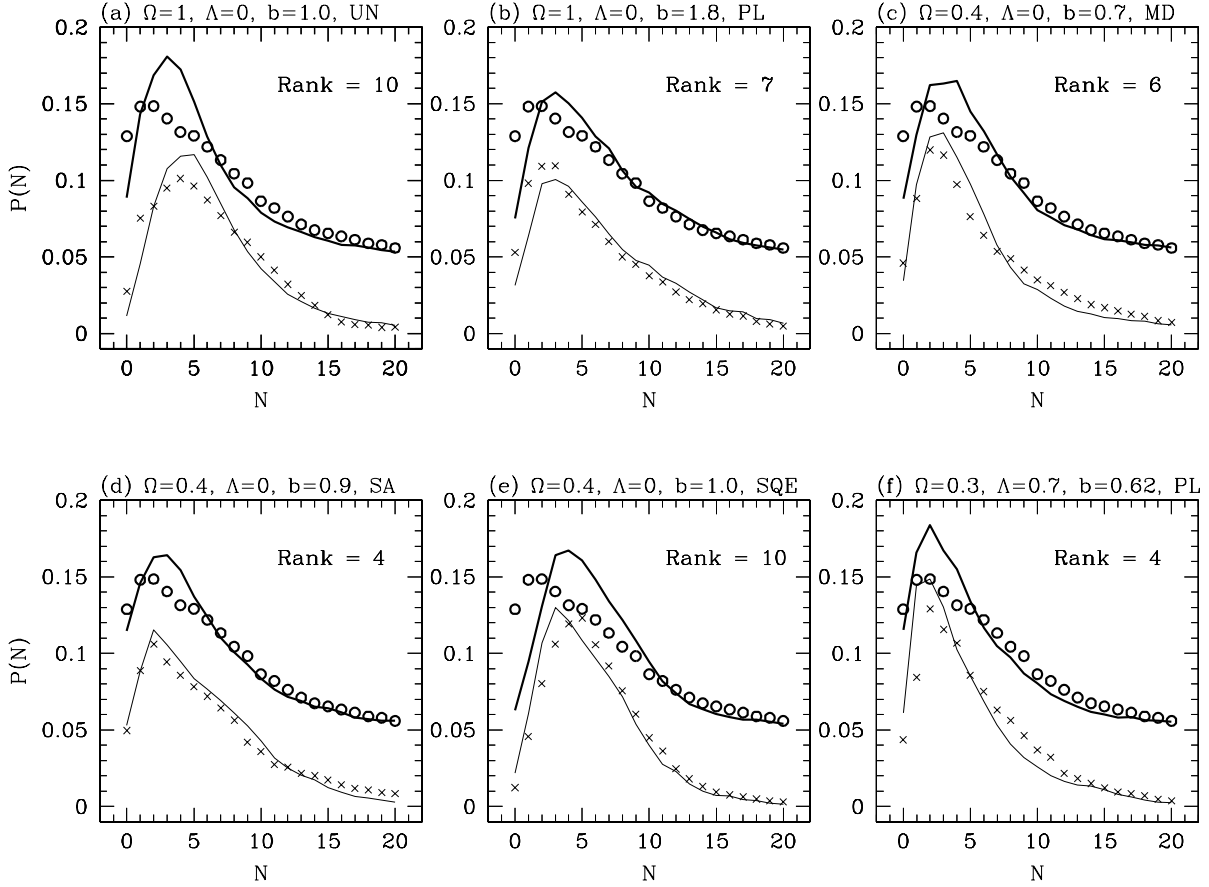}
}
\caption{Distributions of galaxy counts in spheres of radius $8 \hmpc$ 
in the true and reconstructed galaxy distributions.
The crosses and the thin solid line show the counts in cells of the true and 
reconstructed galaxy distributions of the mock catalog ranked $5$
according to the FOM for this statistic (eq.~\ref{eqn:fomcounts}).
The open circles and the thick solid line show the same quantities for the 
PSCz catalog, and are vertically offset by 0.05 for clarity.
The different panels correspond to the galaxy distributions reconstructed 
using the various models listed in Fig.\ 4.
\label{fig:counts}
}
\end{figure}

Figure 9 shows the void probability function (VPF) in the true and
reconstructed galaxy distributions for the six models.
Like the PDF and the count distributions, this statistic is sensitive to 
higher-order correlations in the density field (\cite{white79}; 
\cite{balian89}; \cite{sheth96}), and it can distinguish between biased and 
unbiased galaxy formation models (\cite{little94}).
The probability $P_{0}(R)$, that a randomly placed sphere of 
radius $R$ is devoid of galaxies is a subset of the more general
count distribution statistic $P_{N}(R)$, but here we examine $P_{0}$
at a range of $R$ instead of $P_{N}$ at fixed $R = 8 \hmpc$, as in
Figure 8.
When  computing the VPF, we require that at least $90\%$ of the spherical 
cell's volume lie within the survey region.
We define the FOM for this statistic as
\be
{\rm FOM}_{\rm VPF}  = \sum_{R=0}^{R=\infty} \vert P_{0t}(R) - P_{0r}(R) \vert,
\label{eqn:fomvpf}
\ee
where $P_{0t}(R)$ and $P_{0r}(R)$ are the VPFs of the true and reconstructed 
galaxy distributions, and the sum extends over discrete bins in $R$.
By visual comparison with the mock catalog reconstructions, we expect the 
models E1UNb1.0, E1PLb1.8, O4MDb0.7, and O4SQEb1.0 to have high ranks.
This expectation is confirmed by the ranks of the \pscz reconstruction
listed in each panel.
We also computed the underdensity probability function (UPF) introduced by
Weinberg \& Cole (1992) and found that the different models have similar
ranks for the UPF as for the VPF.
The UPF requires that a sphere be more than $80\%$ below the mean density
rather than completely empty.

\begin{figure}
\centerline{
\epsfxsize=\hsize
\epsfbox[18 144 592 718]{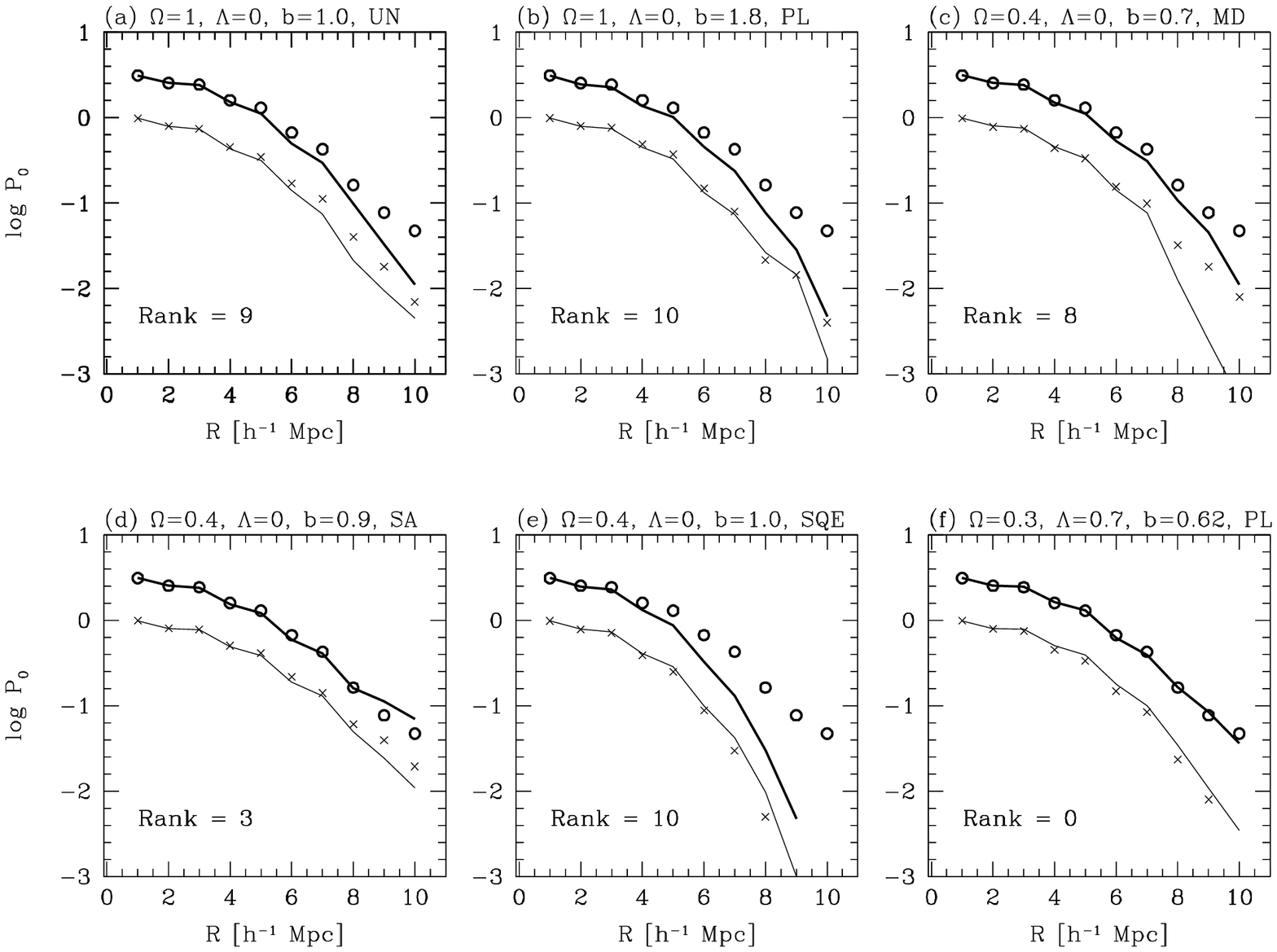}
}
\caption{Void probability functions (VPF) of the true and reconstructed 
galaxy distributions.
The crosses and the thin solid line show the VPFs
of the true and reconstructed galaxy distributions of the mock catalog
ranked $5$ according to the FOM for this statistic (eq.~\ref{eqn:fomvpf}).
The open circles and the thick solid line show the same quantities for the 
PSCz catalog, and are vertically offset by 0.5 for clarity.
The different panels correspond to the galaxy distributions reconstructed 
using the various models listed in Fig.\ 4.
\label{fig:vpf}
}
\end{figure}

Figure 10  shows the distribution of distances to nearest neighbors in the 
true and reconstructed catalogs.
If computed in three dimensions using galaxy redshift distances, this 
statistic would show a spurious peak at neighbor separations corresponding 
to the velocity dispersions of typical galaxy groups.
Therefore, we instead estimate the nearest-neighbor \distrbn from the 
redshift-space \gal distributions using the method suggested by 
Weinberg \& Cole (1992).
For every galaxy at a redshift $z$, we consider all the galaxies that lie
 within a redshift range $\Delta v < 1000\ {\rm km s^{-1}}$ to be its
potential nearest neighbor.
Of these candidate neighbors, we then choose the galaxy that lies closest
to this galaxy in the transverse direction, and we compute
the distribution of this transverse separation $R_{t}$ divided by 
the mean inter-galaxy separation $\bar d$ (i.e., $x_{n} = R_{t}/\bar d$).
This approach biases the estimated neighbor distance, but the bias is the 
same for the \pscz data and the reconstructions.
We define the FOM for this statistic as
\be
{\rm FOM}_{\rm NNBR}  = \sum_{x_{n}=0}^{x_{n}=1} \vert P_{t}(x_{n}) - P_{r}(x_{n}) \vert,
\label{eqn:fomnnbr}
\ee
where $P_{t}(x_{n})$ and $P_{r}(x_{n})$ are the nearest-neighbor distributions  
of the true and reconstructed galaxy distributions.
From the ranks of the \pscz \rec in different panels,
we find that the models E1PLb1.8 and O4SQEb1.0 
have high ranks, while the other models have low ranks.

\begin{figure}
\centerline{
\epsfxsize=\hsize
\epsfbox[18 144 592 718]{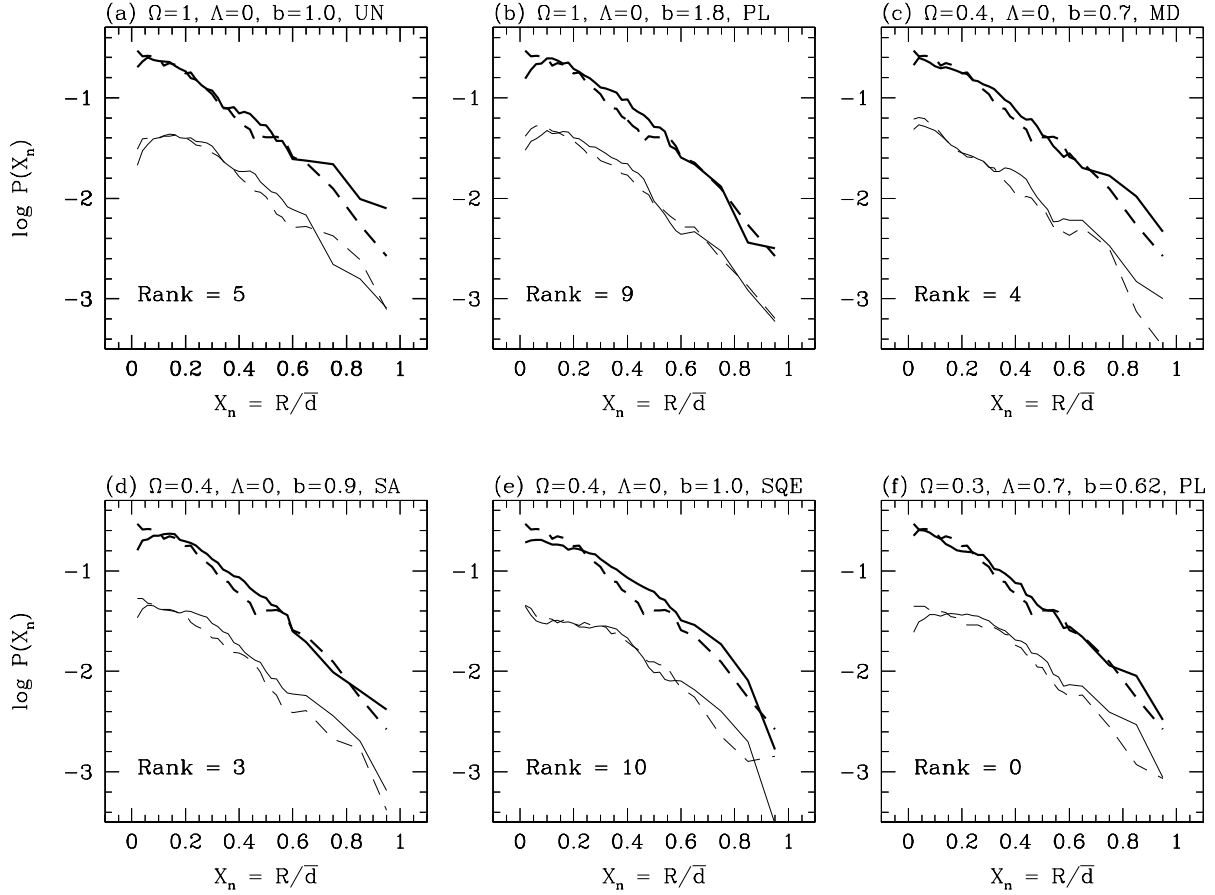}
}
\caption{Distribution of distances to nearest neighbors in the true and 
reconstructed galaxy distributions.
We compute this distribution from the redshift-space locations of galaxies
using the method described in the text.
The thin dashed and solid lines show the nearest neighbor distributions
of the true and reconstructed galaxy distributions of the mock catalog
ranked $5$ according to the FOM for this statistic (eq.~\ref{eqn:fomnnbr}).
The thick dashed and solid lines show the same quantities for the PSCz 
catalog, and are vertically offset by 0.7 for clarity.
The different panels correspond to the galaxy distributions reconstructed 
using the various models listed in Fig.\ 4.
\label{fig:nnbr}
}
\end{figure}

Figure 11 shows the redshift-space correlation functions $\xi(s)$ for the
true and reconstructed catalogs.
We compute the correlation functions using the estimator of Hamilton (1993),
\be
\xi(s) = \frac{N_{DD}N_{RR}}{N_{DR}^{2}} - 1,
\label{eqn:xis}
\ee
where $N_{DD}, N_{DR}$, and $N_{RR}$ are the number of galaxy-galaxy,
galaxy-random, and random-random pairs with a redshift-space separation $s$.
We use a random catalog that has the same geometry and selection function
as the \pscz catalog and contains 50,000 points distributed randomly 
within the survey volume.
We consider only those galaxy pairs that subtend an angle smaller than 
$\alpha_{max} = 60^{\circ}$ at the observer so that the lines of sight 
to both the galaxies in the pair are approximately parallel.
We fit the correlation function in the region $1 \hmpc < s < 15 \hmpc$ 
with a power-law of the form 
$\xi(s) = \left(\frac{s}{s_{0}}\right)^{\gamma}$, where $s_{0}$ is the 
redshift-space correlation length and $\gamma$ is the index of the power-law.
We define the FOM as
\be
{\rm FOM}_{\xi}  = \vert \gamma_{t} - \gamma_{r} \vert,
\label{eqn:fomxis}
\ee
where $\gamma_{t}$ and $\gamma_{r}$ are the slopes of the true and
reconstructed, redshift-space correlation functions.
We find that the model E1UNb1.0 has a high rank, while the other five 
models have low ranks.
The failure of the model E1UNb1.0 is the one expected if we reconstruct a 
low $\Omega_{m}$ universe using a high value of $\Omega_{m}$: 
the high velocity dispersion in an $\Omega_{m}=1$ model
leads to excessive suppression of $\xi(s)$ on small scales, while the
large amplitude of the coherent bulk motions (\cite{kaiser87}) boosts it
on large scales.
Thus, the reconstructed $\xi(s)$ has a shallower slope compared to the
true $\xi(s)$.
We investigated a number of alternative FOM definitions, but we found
(using the mock catalogs rather than the \pscz data set itself)
that the difference in slopes was the most effective measure for
picking out the characteristic signature of excessive redshift space 
distortions.

\begin{figure}
\centerline{
\epsfxsize=\hsize
\epsfbox[18 144 592 718]{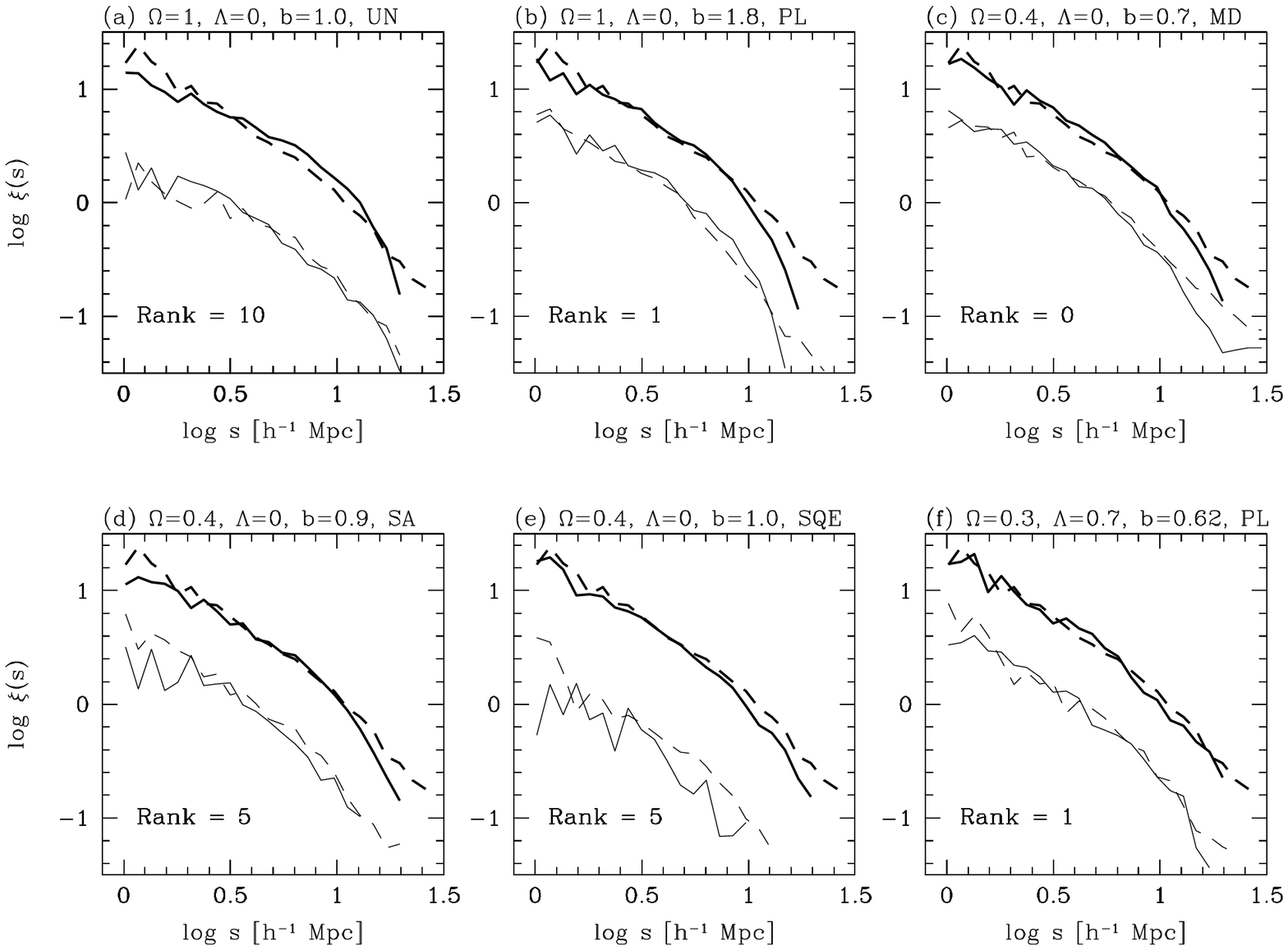}
}
\caption{The redshift-space correlation function, $\xi(s)$, of 
the true and reconstructed galaxy distributions.
The thin dashed and solid lines show $\xi(s)$
of the true and reconstructed galaxy distributions of the mock catalog
ranked $5$ according to the FOM for this statistic (eq.~\ref{eqn:fomxis}).
The thick dashed and solid lines show the same quantities for the PSCz 
catalog, and are vertically offset by 0.5 for clarity.
The different panels correspond to the galaxy distributions reconstructed 
using the various models listed in Fig.\ 4.
\label{fig:xis}
}
\end{figure}

The peculiar velocities of galaxies affect the redshift space clustering 
on both small and large scales, as discussed in \S2.2.
However, the real to redshift space mapping does not affect the galaxy
clustering perpendicular to the line-of-sight.
Figure 12 shows the projected correlation function $w(r_{p})$ (\cite{dp83};
\cite{fisher94}) of the true and reconstructed galaxy distributions, 
computed using an estimator similar to the one defined in 
equation~(\ref{eqn:xis}).
Here, the transverse separation $r_{p}$ is defined by the relation 
$r_{p}^{2} = s^{2} - \pi^{2}$, where $s$ is the true separation in 
redshift-space and $\pi$ is the separation along the line-of-sight between 
the two galaxies in a pair.
We fit a power-law to this function in the range $1 \hmpc < r_{p} < 15 \hmpc$,
and define a FOM similar to that defined by equation~(\ref{eqn:fomxis}).
We find that the model O4SQEb1.0 has a high rank,
while the other five models have low ranks.

\begin{figure}
\centerline{
\epsfxsize=\hsize
\epsfbox[18 144 592 718]{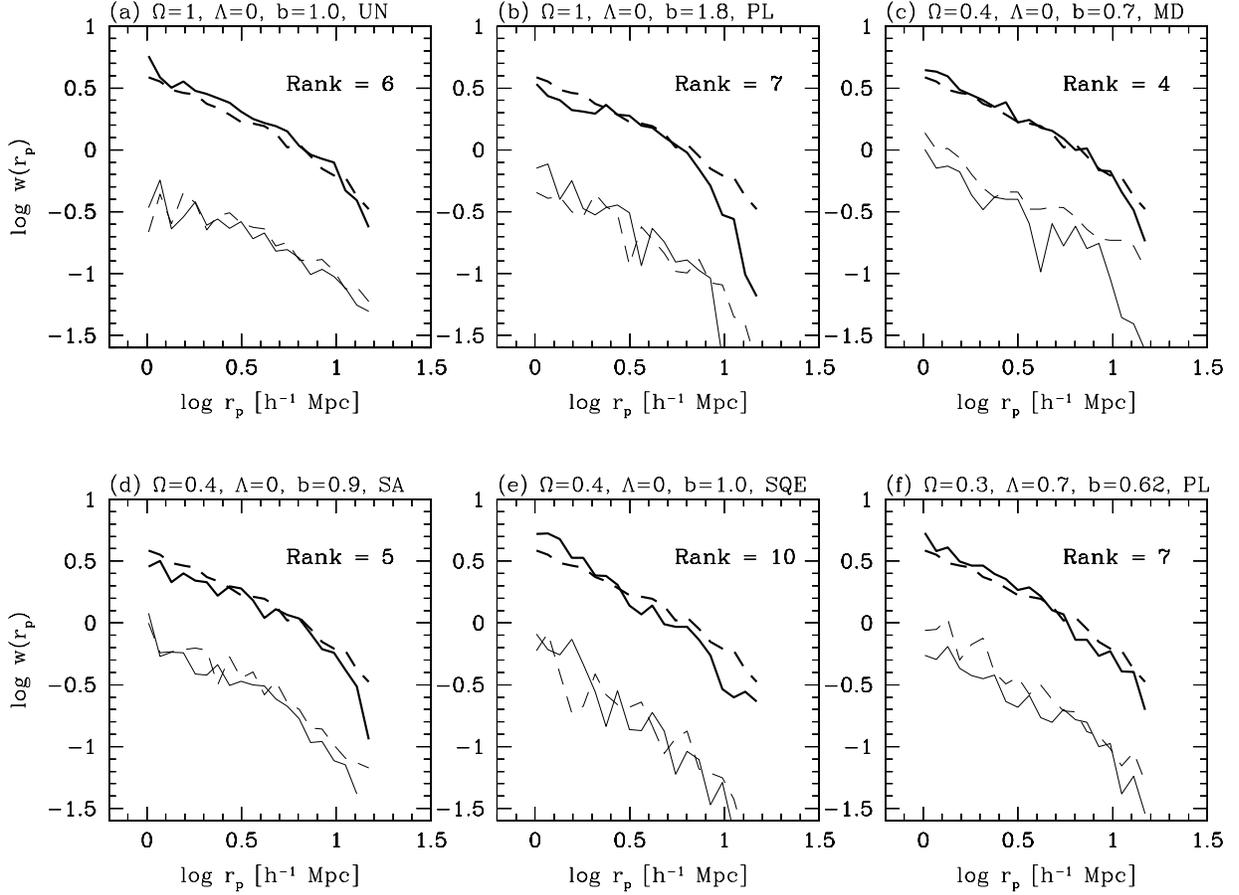}
}
\caption{The projected correlation function, $w(r_{p})$, of 
the true and reconstructed galaxy distributions.
The thin dashed and solid lines show $w(r_{p})$
of the true and reconstructed galaxy distributions of the mock catalog
ranked $5$ according to the FOM for this statistic (eq.~\ref{eqn:fomxis}).
The thick dashed and solid lines show the same quantities for the PSCz 
catalog, and are vertically offset by 0.5 for clarity.
The different panels correspond to the galaxy distributions reconstructed 
using the various models listed in Fig.\ 4.
\label{fig:wrp}
}
\end{figure}

\section{Evaluation of Models}

We also reconstructed the \pscz catalog and the corresponding mock catalogs 
for the remaining set of nine models described briefly in \S3.2.
For all 15 models, we measured all the statistics described in the last 
section.
We then ranked these models using the FOM corresponding to each statistic, 
in the manner described in \S4.
Table 2 lists the ranks of the \pscz \rec with respect to the mock catalogs
for our full suite of 15 models.

Table 2 is the complete quantitative summary of the results of our \rec 
analysis of the \pscz catalog.
A low rank for any statistic indicates that the model reproduces that 
statistical property of the \pscz catalog as well as, or better than, 
it reproduces that property for most of the mock catalogs corresponding 
to that model.
On the other hand, a high rank (close to 10) for any statistic indicates 
that the model does not reproduce that property of the \pscz catalog as 
accurately as would be expected (based on the mock catalogs) if the model
were a correct representation of the real universe.
Computational practicality limits us to ten mock catalogs for each of our
models, so even if the \pscz \rec has a rank of 10 for a particular
statistic, we can only conclude that the model fails that statistical test
at the $\sim 90\%$ confidence level.
If we were to reconstruct 100 mock catalogs for that model,
we would expect the \pscz \rec to be worse than at least $\sim 90$ of the mock
catalogs (unless we happened to be unusually lucky in the ten that we did
reconstruct), but we do not know whether it would be worse than 95, or 99, or
all 100, since we have not been able to probe the tails of the \rec
error distribution.

Two issues complicate the interpretation of Table 2.
First is the fact that we have considered many different statistical tests
and therefore given the \pscz reconstruction many ``chances to fail''.
As a result, a single rank of 10 does not necessarily imply a failure of 
the model; if the nine statistics were entirely independent of each
other (which they are not), we would expect a typical mock catalog to have 
one rank of 10, and a significant fraction to have more than one
rank of 10.
In order not to be misled, we must compare the ranks of the \pscz \rec 
with the ranks of  the mock catalog reconstructions even when we draw 
general inferences from Table 2, as we do below.

The second complication is that the statistical measures are not all
independent of each other, since the clustering properties that they
quantify are in some cases closely related.
Fortunately, we can use the mock catalogs themselves to understand the 
correlations between the different statistics.
Using all 150 mock catalogs, we computed the covariance matrix of the
ranks of the nine statistics, and we also computed the distribution of 
mock catalog ranks for each statistic conditioned on the catalog having
a rank of 10 for one of the other statistics.
Both analyses led to the same conclusion: the nine statistics fall into
five groups, and ranks within each group are correlated but ranks in one
group are essentially uncorrelated with ranks in another group.
The five groups are:
(1) the correlation coefficient $(r)$ and the Fourier difference 
statistic $D(k)$, (2) the PDF of the smoothed galaxy density field, 
(3) the counts in spheres of  radii $3 \hmpc$ and $8 \hmpc$ and the void 
probability function, (4) the nearest  neighbor distribution, and (5) 
the two correlation functions $\xi(s)$ and $w(r_{p})$.
The statistics in the first and third groups are strongly correlated 
amongst themselves, while the statistics in the fifth group (the two 
correlation functions) are only moderately correlated.

As an overall quantitative measure of the success of a \pscz \rec relative
to the expectation based on mock catalogs, we list the weighted mean rank of
the \pscz \rec in column 11 of Table 2.
We weight the rank of each statistic inversely by the number of statistics
in its correlated group, so each of the five independent groups contributes
equally to this mean rank.
The mean weighted rank of mock catalogs computed in this manner is
$5.0$, so a \pscz \rec with mean rank greater than $5.0$ is less accurate
than a \rec of a typical mock catalog, and vice versa.

Two of the models listed in Table 2 fail the \rec test unambiguously.
The \pscz \rec for model O4UNb1.0 has a rank of 10 in each of three
independent groups of statistics, the PDF, the counts/VPF, and the 
nearest neighbor distribution.
The worst of the ten mock catalog reconstructions of this model has
two independent ranks of 10, while the next worst has one rank of 10 
and two ranks of 9.
The O4SQEb1.0 model fails even more clearly.
The \pscz \rec of this model has a rank of 10 in four of the five independent
groups of statistics, while the worst mock catalog \rec for this model has one
rank of 10 and one rank of 9.
For both models, the weighted mean rank of the \pscz \rec is higher than that 
of any of the model's ten mock catalog reconstructions.
Remarkably, the O4SAb0.9 model, which has nearly the same cosmological
parameters as these two failed models but a different form of the biasing
relation, produces one of the most successful \pscz reconstructions.
We return to this point in \S6.

The other model that fares especially poorly in Table 2 is E1UNb1.0,
which has two independent ranks of 10 and a rank of 9 in third
independent group.
One of the mock catalog reconstructions of this model actually 
performs worse, with three independent ranks of 10 and a fourth
independent rank of 9, and this mock catalog has a weighted mean
rank of $7.7$, identical to the \pscz mean rank of $7.7$.
In a purely statistical sense, therefore, we cannot rule out this
model as clearly as we can rule out O4UNb1.0 and O4SQEb1.0.
However, as already noted in our discussion of Figure 11a, the E1UNb1.0
\rec of \pscz fails in exactly the manner expected if we reconstruct  a
low $\Omega_{m}$ universe with an $\Omega_{m}=1$ model of similar 
$\sigma_{8m}$: the high peculiar velocities in the $\Omega_{m}=1$
\rec suppress $\xi(s)$ on small scales and boost it on large scales,
making the reconstructed slope of $\xi(s)$ too shallow.
The \pscz \rec has a rank of 10 for $\xi(s)$ but only 6 for $w(r_{p})$,
so it is clear that this failure is caused by the reconstruction's
excessive peculiar velocities, not by a problem in the real space
clustering.
None of the mock catalog reconstructions of this model, including the
one that has more high ranks than the \pscz reconstruction, fails in this
characteristic manner.

Based on these considerations, we classify the models O4UNb1.0, O4SQEb1.0
and E1UNb1.0 as Rejected according to our analysis, and we indicate this
classification by an R in column 12 of Table 2.
These three \pscz reconstructions have the highest mean ranks among the
15 models, 7.8, 7.8 and 7.7, respectively.
We classify the remaining 12 models as Accepted (indicated by an A in 
column 12), since in each of case there is at least one of the model's
mock catalogs that has more independent ranks of 9 or 10 than the model's
\pscz reconstruction.
However, within this class of Accepted models, there is a wide range in the
relative accuracy of the \pscz reconstruction.
The models E1PLb1.3, E1PLb1.8, and L2PLb0.77 are all Accepted on the
basis of a single mock catalog that is
reconstructed worse than the \pscz catalog, while the models E1PLMDb1.3 
and O3PLb1.4 are Accepted on the basis of two mock catalogs that are 
reconstructed worse than the \pscz catalog.
These five  models have the highest mean ranks among the 12 Accepted
models.
In each of the remaining seven Accepted models, the \pscz is reconstructed
better than at least three mock catalogs, and these models have 
correspondingly smaller average ranks for their \pscz reconstructions.

\begin{table}
\caption{Rank of the PSCz reconstruction with respect to the mock catalogs, 
for the full suite of 15 models.
Column 1 gives the name of the model.
Column 2 lists the rank of the PSCz reconstruction according to the 
correlation between the true and reconstructed smoothed  galaxy density 
fields $(r)$, while the number enclosed in brackets shows the absolute value 
of the correlation coefficient.
For each model, columns 3 to 10 list the rank of the PSCz reconstruction with
respect to the corresponding mock catalogs, for the following statistical 
properties:
the Fourier difference statistic $D(k)$,
the PDF of the smoothed galaxy density field, 
the galaxy counts in spheres of radii $3 \hmpc$ and $8 \hmpc$,
the void probability function (VPF), 
the nearest neighbor distribution (NNBR),
the redshift-space correlation function $\xi(s)$,
and the projected correlation function $w(r_{p})$.
Column 11 lists the weighted mean rank for the \pscz reconstruction computed
using the procedure described in the text.
Column 12 lists our qualitative classification, Accepted(A) or Rejected(R),
as described in the text.
In all cases, the rank of the \pscz is the number of mock catalogs
(out of 10) that are reconstructed more accurately than the model 
reconstructs PSCz, according to the FOM for the corresponding statistic.
}
\medskip
\begin{tabular}{*{13}{c|}}
\tableline\tableline
& & & & \multicolumn{2}{c|}{Counts} & & & & & & \\ \cline{5-6}
Name & r & D(k) & \multicolumn{1}{c|}{PDF} & $3h^{-1}$ & $8h^{-1}$ & VPF & NNBR & $\xi(s)$ & $w(r_{p})$ & $\left<{\rm Rank}\right>$ & Status \\ 
& & & & Mpc & Mpc & & & & & & \\ 
\tableline
E1UNb1.0     & 7(0.73) & 9  & 8  & 9  & 10 & 9  & 5   & 10 & 6 & 7.7 & R \\
E1PLb1.3     & 7(0.77) & 6  & 7  & 10 & 10 & 10 & 7   & 5  & 6 & 7.2 & A \\
E1PLMDb1.3   & 4(0.75) & 5  & 4  & 9  & 8  & 8  & 6   & 4  & 7 & 5.7 & A \\
E1THb1.3     & 4(0.74) & 4  & 6  & 10 & 9  & 7  & 7   & 0  & 4 & 5.5 & A \\
E1PLb1.8     & 0(0.86) & 2  & 8  & 8  & 7  & 10 & 9   & 1  & 7 & 6.1 & A \\
O2PLb0.5     & 1(0.57) & 2  & 2  & 0  & 3  & 0  & 3   & 0  & 7 & 2.2 & A \\
O3PLb1.4     & 1(0.85) & 2  & 6  & 10 & 9  & 10 & 10  & 0  & 5 & 5.9 & A \\
O4UNb1.0     & 5(0.80) & 8  & 10 & 10 & 10 & 9  & 10  & 3  & 3 & 7.8 & R \\
O4MDb0.7     & 1(0.83) & 3  & 9  & 8  & 6  & 8  & 4   & 0  & 4 & 4.9 & A \\
O4SAb0.9     & 4(0.79) & 5  & 3  & 7  & 4  & 3  & 3   & 5  & 5 & 4.0 & A \\
O4SQEb1.0    & 1(0.76) & 2  & 10  & 10  & 10  & 10  & 10   & 5  & 10 & 7.8 & R \\
L2PLb0.77    & 1(0.75) & 1  & 9  & 7  & 10 & 8  & 9   & 4  & 6 & 6.5 & A \\
L3PLb0.62    & 3(0.63) & 3  & 0  & 4  & 4  & 0  & 0   & 1  & 7 & 1.9 & A \\
L4PLb0.64    & 3(0.62) & 4  & 7  & 3  & 7  & 3  & 8   & 0  & 8 & 5.4 & A \\
L5PLb0.54    & 6(0.43) & 7  & 0  & 2  & 3  & 0  & 1   & 10 & 2 & 3.0 & A \\
\tableline
\end{tabular}
\label{table:tab1}

\end{table}

\section{Discussion}

We have reconstructed the IRAS galaxy distribution in our cosmological 
neighborhood, within a spherical region of radius $50 \hmpc$ centered on the
Local Group.
We have tested 15 different models, each consisting of a set of assumptions 
about the values of $\Omega_{m}$ and $\Omega_{\Lambda}$, the value of the 
bias factor $b$, and the nature of the biasing relation between IRAS galaxies 
and the underlying mass.
For every model, we have quantified the accuracy of the \pscz reconstruction
relative to the expectation based on mock \pscz catalogs, and we have used 
this result to classify the model as Accepted or Rejected.
The Rejected models are unlikely to be the correct models for 
structure formation in the real universe, while for the Accepted models the
\pscz \rec is more accurate than at least one of the model's ten mock
catalog reconstructions.
We have computed mean weighted  ranks (Table 2, column 11) as an overall
quantitative measure of the accuracy of a model's \pscz \rec relative
to the expectation from mock catalogs.
We now examine these results in detail to see what general 
conclusions we can derive regarding the allowed ranges of cosmological 
and galaxy formation parameters.

Figure 13 shows the locations of all $15$ models in the 
$\Omega_{m} - \sigma_{8m}$ plane.
The 12 distinct points correspond to the $15$ different models because 
there are sets of models with identical values of $\Omega_{m}$ and 
$\sigma_{8m}$ (and hence $b$) but with different biasing schemes.
Thus, for example, the two models O4UNb1.0 and O4SQEb1.0 are indistinguishable
in this plane, as are the three models E1PLb1.3,  E1PLMDb1.3, and E1THb1.3. 
The circles show the 12 Accepted models, and the triangles show the 
three Rejected models.
For the Accepted models, the radius of the circles is proportional to 
$(10 - \left< {\rm Rank} \right>)$, where $\left< {\rm Rank} \right>$ is
the mean rank for the \pscz \rec of a model.
Hence, larger circles show models that are more successful in reconstructing
the \pscz catalog.
The shaded region shows the observed rms fluctuation of the IRAS galaxy 
distribution, $\sigma_{8g}({\rm IRAS}) = 0.69 \pm 0.04$ (\cite{fisher94}).

We plot four different constraints in the $\Omega_{m} - \sigma_{8m}$ plane
that are obtained using independent techniques.
The solid line in Figure 13 shows the constraint 
$\sigma_{8m} \Omega_{m}^{0.6} = 0.55$, 
required to reproduce the observed masses and abundances of rich clusters 
of galaxies at the present epoch (\cite{wef93}; \cite{eke96}; \cite{viana96}).
The dotted line shows the constraint $\sigma_{8m}\Omega_{m}^{0.6} = 0.85$, 
which is implied by the power spectrum of mass density fluctuations estimated
from the peculiar velocities of galaxies in the SFI catalog 
(\cite{freudling99}).
The remaining two constraints arise from comparing the IRAS galaxy distribution
with the peculiar velocities of galaxies.
In linear perturbation theory, the mass continuity equation takes the form
(\cite{lss80})
\be
\divv = - f H_{0}\delta_{m},
\label{eqn:continuity}
\ee
where $f \approx \Omega_{m}^{0.6}$.
If we assume that $\delta_{g} = b_{\delta}\delta_{m}$ (the linear bias model),
equation~(\ref{eqn:continuity}) becomes 
\be
\divv = -\beta H_{0}\delta_{g},
\label{eqn:continuity2}
\ee
where $\beta = \Omega_{m}^{0.6}/b_{\delta}$.
The short dashed line shows the constraint $\beta_{\rm IRAS} = 0.5$ obtained 
by the VELMOD method, which derives a maximum likelihood estimate of 
$\beta_{\rm IRAS}$ by comparing the peculiar velocities of galaxies in the 
Mark III catalog with their radial velocities predicted from the IRAS 1.2 Jy 
redshift catalog, using 
equation~(\ref{eqn:continuity2}) (\cite{willick97a}; \cite{willick98}).
This value of $\beta_{\rm IRAS}$ is also obtained from an analysis of the 
anisotropy of the redshift-space power spectrum of IRAS galaxies 
(Cole, Fisher, \& Weinberg 1995), and from a comparison of the
spherical harmonics of the peculiar velocity field derived from the 
Mark III catalog with the spherical harmonics of the gravity field derived 
from the IRAS 1.2 Jy survey 
(Davis, Nusser, \& Willick 1996; see Strauss \& Willick 1995 and
Hamilton 1998 for reviews of other estimates of $\beta_{\rm IRAS}$).
We convert an estimate of $\beta_{\rm IRAS}$ into a constraint on 
$\sigma_{8m}\Omega_{m}^{0.6}$ using the relation
\be
\sigma_{8m}\Omega_{m}^{0.6} = \beta_{\rm IRAS} \sigma_{8g},
\label{eqn:beta2sigom}
\ee
where we have assumed that $b_{\delta} = b =  \sigma_{8g}/\sigma_{8m}$.
The long dashed line shows the constraint $\beta_{\rm IRAS} = 0.9$ obtained by 
the POTENT method, which measures $\beta_{\rm IRAS}$ as the slope of the 
regression between the observed galaxy density field from the IRAS 1.2 Jy 
redshift catalog and the mass density field derived from the peculiar 
velocities of galaxies in the Mark III catalog, using a modified version of 
equation~(\ref{eqn:continuity}) (\cite{sigad98}).
Each of these four constraints has a quoted uncertainty of about $10\% - 20\%$,
implying that all of them cannot be consistent with one another.

\begin{figure}
\centerline{
\epsfxsize=\hsize
\epsfbox[18 220 592 718]{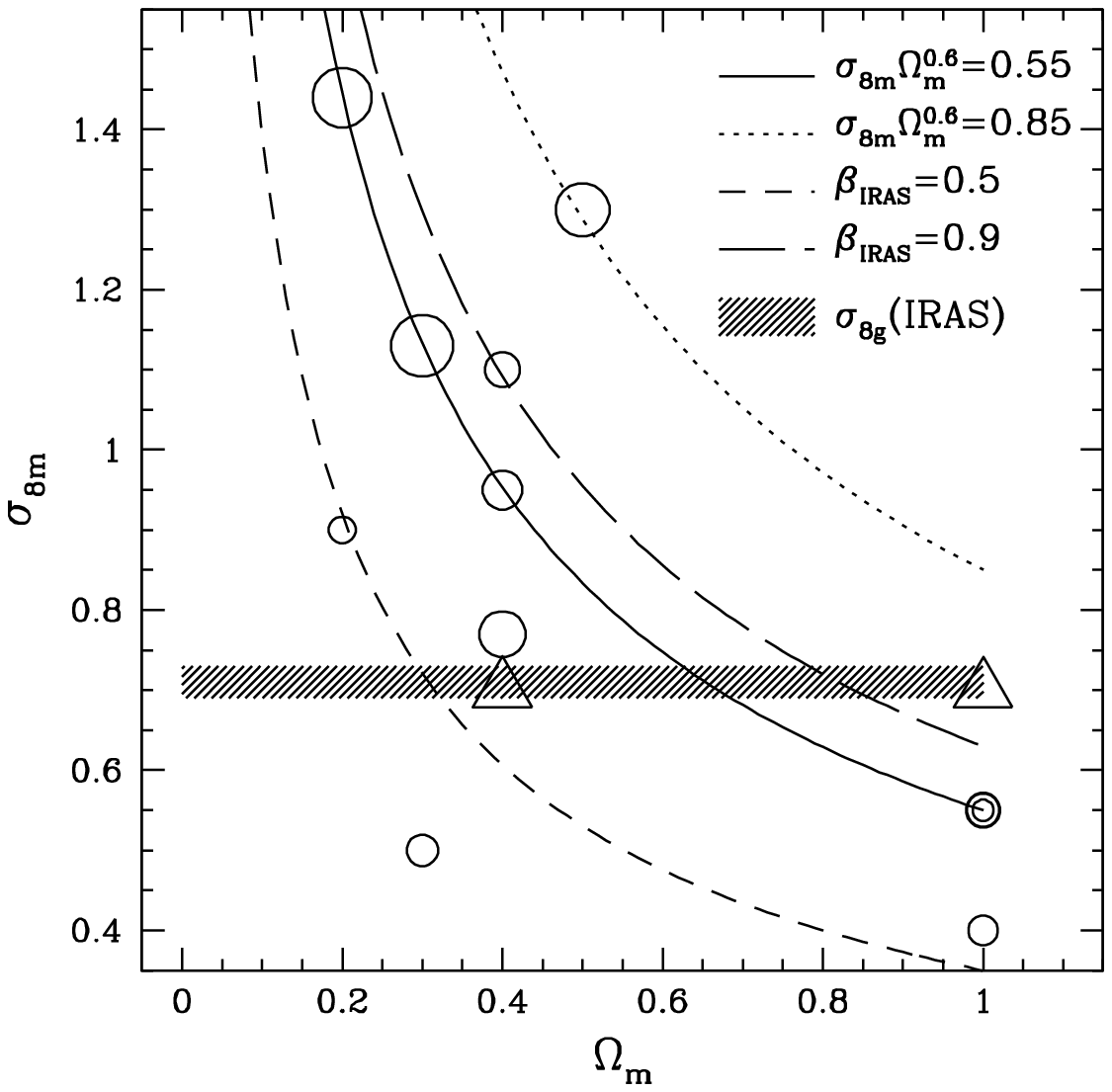}
}
\caption{Points show the locations of the 15 models in the 
$\Omega_{m}-\sigma_{8m}$ plane, classified as Accepted (circles), 
or Rejected (triangles).
The radius of the circles is proportional to 
$(10 - \left< {\rm Rank} \right>)$, 
where $\left< {\rm Rank} \right>$ is the mean rank for the \pscz 
\rec of a model; larger circles represent more successful reconstructions.
The lines show constraints in this plane derived from four different
techniques: 
(1) $\sigma_{8m}\Omega_{m}^{0.6} = 0.55$ from the abundance of clusters 
at $z=0$ (solid line), (2) $\sigma_{8m}\Omega_{m}^{0.6} = 0.85$ from a maximum
likelihood estimate of the mass power spectrum using peculiar velocity data 
(dotted line), 
(3) $\beta_{\rm IRAS} = 0.5$ from the VELMOD analysis of the Mark III catalog 
of peculiar 
velocities and the IRAS 1.2 Jy galaxy redshift catalog (short dashed line), and
(4) $\beta_{\rm IRAS} = 0.9$ from the POTENT analysis of the Mark III catalog of peculiar 
velocities and the IRAS 1.2 Jy galaxy redshift catalog (long dashed line).
The shaded region shows the observed fluctuation amplitude of the IRAS galaxy 
distribution, $\sigma_{8g}({\rm IRAS}) = 0.69 \pm 0.04$.
\label{fig:constr}
}
\end{figure}

Based on the ranks of the 15 models in Table 2, and their locations 
in Figure 13, we arrive at the following conclusions.
\begin{description}
\item[{(1)}: ] Our successful reconstructions of the \pscz catalog, at least 
for some plausible assumptions about the value of $\Omega_{m}$ and the bias 
between IRAS galaxies and mass, lends support to the hypothesis that LSS 
originated in the gravitational instability of small amplitude, Gaussian 
primordial mass density fluctuations.
While this success does not, by itself, rule out non-Gaussian models for 
primordial fluctuations, it strengthens the viability of Gaussian models.
Models whose initial conditions have substantially non-Gaussian PDFs
generally predict quite different properties for LSS 
(\cite{moscardini91}; \cite{wc92}).
\item[{(2)}: ] Unbiased models in which IRAS galaxies trace mass
are rejected, for both $\Omega_{m} = 0.4$ and $\Omega_{m} = 1$.
From Table 2 and the discussion in \S5, it is clear that the models E1UNb1.0 
and O4UNb1.0 are both clearly rejected by the \rec analysis of the \pscz 
catalog.
Figure 11(a) shows that the model E1UNb1.0 fails in the manner expected
if we reconstruct the redshift-space galaxy distribution in a low 
$\Omega_{m}$ universe using, erroneously, a high value of $\Omega_{m}$.
The high velocity dispersion of clusters in an $\Omega_{m}=1$ model suppresses 
the small scale correlations in redshift space, while the large scale bulk 
flows, whose amplitude is proportional to $\Omega_{m}^{0.6}$, enhances the 
correlations on large scales.
Therefore, the reconstructed $\xi(s)$ has a shallower slope compared to the
true $\xi(s)$, although the rms fluctuations (in redshift space) of the 
two galaxy distributions are similar, by construction.
\item[{(3)}: ] Of the five models with $\Omega_{m} = 1$, 
E1THb1.3 is the only Accepted model that reconstructs the \pscz catalog as 
well as its own typical mock catalog.
While the model E1UNb1.0 is clearly Rejected, the models E1PLb1.3 and E1PLb1.8
are Accepted because one mock catalog in each of these models is reconstructed
worse than the \pscz catalog, and the model E1PLMDb1.3 is Accepted
because there are two mock catalogs that are reconstructed worse than
the \pscz catalog.
Thus, although four of the five models with $\Omega_{m}=1$ are Accepted,
most of them are only moderately successful in reconstructing the
\pscz catalog.
\item[{(4)}: ] In Figure 13, there are five models  that lie on, or close to, 
the constraint $\beta_{\rm IRAS}  = 0.5$.
Of these, the models O4UNb1.0 and O4SQEb1.0 are clearly rejected, 
the models E1Plb1.8 and L2PLb0.77 are Accepted because there is one mock 
catalog that is reconstructed worse than the \pscz catalog, while the 
model O3PLb1.4 is Accepted because there are two mock catalogs that are 
reconstructed worse than the \pscz catalog.
All these models are at best only moderately successful in reconstructing
the \pscz catalog.
This leads us to conclude that the low-normalization constraint 
$\beta_{\rm IRAS}  = 0.5$ 
(corresponding to $\sigma_{8m}\Omega_{m}^{0.6} \approx 0.35$),
inferred from the simplest interpretation of the VELMOD (\cite{willick98})
and redshift-space distortion (\cite{cfw95}) analyses of the $1.2$Jy
redshift survey, is only marginally successful in reconstructing the
\pscz catalog.
Here we have assumed that $b_{\delta} = b = \sigma_{8g}/\sigma_{8m}$ to 
convert an estimate of $\beta_{\rm IRAS}$ into a constraint on 
$\sigma_{8m}\Omega_{m}^{0.6}$.
While this relation is valid in a deterministic, linear 
bias model, in the case of a more general biasing relation between galaxy and 
mass distributions, the relation between $b$ and $b_{\delta}$ also includes 
terms arising from the non-linearity and the stochasticity of the biasing 
relation (\cite{dekel99}).
We are currently investigating the extent to which the estimates of
$\beta_{\rm IRAS}$ using different techniques, including POTENT, VELMOD,
and the anisotropy of the redshift-space power spectrum, are sensitive
to the details of the biasing scheme
(Berlind, Narayanan, \& Weinberg, in preparation).
\item[{(5)}: ] There are seven models including E1THb1.3, O2PLb0.5, 
O4MDb0.7, O4SAb0.9, L3PLb0.62, L4PLb0.64, and L5PLb0.54, in which the \pscz 
catalog is reconstructed better than at least three mock catalogs 
corresponding to that model.
These models are the most successful in reconstructing the properties
of the galaxy distribution in the \pscz catalog.
Except for the model E1THb1.3, all these models have $\Omega_{m} < 1$
and require that $\sigma_{8m} > \sigma_{8g}$ (hence $b < 1$), 
i.e., that IRAS galaxies be antibiased with respect to the mass distribution 
on a scale of $8 \hmpc$.
However, we are unable to pin down the bias factor more precisely:
the model O4SAb0.9, with a small antibias, and the model O2PLb0.5,
with a large antibias, both reconstruct the \pscz catalog very well.
Most of the successful models require that $\beta_{\rm IRAS} \ge 0.8$
(except the model O4SAb0.9, which requires $\beta_{\rm IRAS} \approx 0.7$)
\item[{(6)}: ] The model O4SAb0.9, in which IRAS galaxies are related 
to the mass \distrbn according to the predictions of the semi-analytic galaxy
formation model, reconstructs the \pscz catalog very well.
This accurate reconstruction of the \pscz catalog is a non-trivial success 
of the semi-analytic model, since the models O4UNb1.0 and
O4SQEb1.0, with quite similar values of $\Omega_{m}$ and $\sigma_{8m}$ but
with different biasing relations, are both clearly rejected.
This sensitivity of the \rec to the {\it nature} of the biasing relation 
demonstrates that the \rec analysis of a galaxy redshift survey can
distinguish between different {\it bias models}, and not just between 
different values of the bias factor $b$.
\end{description}

Conclusions $(2)-(6)$ are based on reconstructing the 
\pscz catalog under the general assumptions that the primordial mass density 
fluctuations form a Gaussian random field and that the bias between IRAS 
galaxies and mass can be characterized by a local, monotonic function.
While the assumption of Gaussian initial fluctuations enables us to 
constrain the nature of the biasing relation, we could also use \rec
analysis in a complementary mode, i.e., to test the level of 
non-Gaussianity of the initial fluctuations, given our current state of 
knowledge of the galaxy formation process.
In this regard, the successful \rec for a model (O4SAb0.9) based on a
physically motivated theory of galaxy formation and cosmological parameter
values supported by independent constraints supports the standard hypothesis
that primordial fluctuations were not far from Gaussian.

There are several natural directions for extending this analysis using
observational data.
In this paper, we have compared the properties of the \rec to the input 
\pscz galaxy distribution in redshift space alone.
However, every model \rec predicts both the real-space galaxy 
distribution and the fully non-linear peculiar velocity field at every 
point within the \rec volume.
We can then compare the velocity field predicted for any model with the 
observed peculiar velocities of galaxies in, say, the Mark III catalog 
(\cite{willick97b}) or the SFI catalog (\cite{giovanelli97}).
Such a comparison will be more accurate than a comparison involving the
velocity field predicted using linear theory.
The amplitude and the non-linear components of the velocity field 
serve as good diagnostics of the allowed values of $\Omega_{m}$ and 
$\sigma_{8m}$ (\cite{nw98}).
We can correct for inhomogeneous Malmquist bias when comparing the
observed density and velocity fields by using the reconstructed 
line-of-sight density  and velocity distributions to every galaxy.
Alternatively, we can circumvent the effects of Malmquist bias
by working directly in redshift space itself (\cite{strauss95}).

In order to understand the galaxy formation process, it is necessary to 
study the relative bias between different types of galaxies as well as the
absolute bias between the galaxy population as a whole and the underlying mass.
For example, it is now well known that optical galaxies are more strongly
clustered than IRAS galaxies (\cite{lahav90}; \cite{strauss92}; 
\cite{saunders92}; \cite{pd94}; \cite{willmer98}; \cite{baker98}).
Reconstruction analysis of the galaxy distribution in the Optical Redshift 
Survey (Santiago et al.\ 1995, 1996), using a set of models similar to the ones
discussed in this paper, will give us independent constraints on 
$\Omega_{m}$ and $\sigma_{8m}$ and enable us to test whether optical 
galaxies trace the underlying mass.
Since the Optical Redshift Survey and \pscz probe similar regions of space,
the initial conditions derived from \rec of the two data should be 
consistent with each other, and it should be possible to reproduce one
catalog beginning with the initial conditions derived from the other
by changing only the biasing model used to select galaxies from the 
evolved mass distribution.

Reconstruction analysis is thus a powerful tool to constrain the ranges 
of allowed values of cosmological parameters and the details of the galaxy 
formation process.
For example, we can discriminate between models with low and high
values of $\Omega_{m}$, and between models with different values of
$\sigma_{8m}$ (hence, different values of $b$).
However, if the cosmological parameters and the mass power
spectrum can be determined precisely using other constraints, such as
Type Ia supernovae, cosmic microwave background anisotropies,
the Lyman-$\alpha$ forest, or  weak lensing, then 
\rec analysis can focus on deriving the biasing relation between the
different galaxy distributions and the underlying mass distribution.
Knowledge of these relations will, in turn, provide  strong tests of
numerical and semi-analytic models for galaxy formation.

\acknowledgments

VKN and DHW were supported by NSF Grant AST-9616822. 
VKN also acknowledges support by the Presidential Fellowship from the
graduate school of The Ohio State University.
We thank Michael Hudson for helpful suggestions.


\begin{thebibliography}{}

\bibitem[Abell 1958]{abell58}
Abell, G. O. 1958, \apjs, 50, 241

\bibitem[Babul \& Postman 1990]{babul90}
Babul, A., \& Postman, M. 1990, \apj, 359, 280

\bibitem[Bagla \& Padmanabhan 1997]{bagla97}
Bagla, J.S., \& Padmanabhan, T. 1997, \mnras, 286, 1023

\bibitem[Baker et al. 1998]{baker98}
Baker, J.E., Davis, M., Strauss, M.A., Lahav, O., \& Santiago, B.X 1998, \apj, 508, 6

\bibitem[Balian \& Schaeffer 1989]{balian89}
Balian, R., \& Schaeffer, R. \aap, 220, 1

\bibitem[Bardeen et al. 1983]{bardeen83}
Bardeen, J.M., Steinhardt, P.J.,  \& Turner, M.S. 1983, Phys Rev D, 28, 679

\bibitem[Beacom et al. 1991]{beacom91}
Beacom, J.F., Dominik, K.G., Melott, A.L., Perkins, S.P., \& Shandarin, S.F. 1991, \apj, 372, 351

\bibitem[Benson et al. 1999]{benson99}
Benson, A.J., Cole, S., Frenk, C.S., Baugh, C.M., \& Lacey, C.G 1999, \mnras, submitted, (astro-ph/9903343)

\bibitem[Bertschinger 1987]{bertschinger87}
Bertschinger, E., 1987, \apj, 323, L103

\bibitem[Branchini et al.  1999]{branchini99}
Branchini, E., et al.\ 1999, \mnras, in press

\bibitem[Brown \& Peebles 1987]{brown87}
Brown, M.E., \& Peebles, P.J.E., 1987, \apj, 317, 588

\bibitem[Canavezes et al.\ 1998]{canavezes98}
Canavezes, A., et al.\ 1998, \mnras, 297, 777

\bibitem[Cen \& Ostriker 1992]{co92} 
Cen, R., \& Ostriker, J. P. 1992, \apj, 399, L113

\bibitem[Cen \& Ostriker 1993]{co93} 
Cen, R., \& Ostriker, J. P. 1993, \apj, 417, 415

\bibitem[Cole et al.\ 1994]{cole94}
Cole, S., Aragon-Salamanca, A., Frenk, C.S., Navarro, J.F., \& Zepf, S.E. 1994, \mnras, 271, 781

\bibitem[Cole et al.\ 1995]{cfw95} 
Cole, S., Fisher, K.B., \& Weinberg, D.H 1995, \mnras, 275 515

\bibitem[Cole et al.\ 1997]{cwfr97}
Cole, S., Weinberg, D. H., Frenk, C. S., \& Ratra, B. 1997, \mnras, 289, 37 

\bibitem[Cole et al.\ 1998]{cole98}
Cole, S., Hatton, S., Weinberg, D. H., \& Frenk, C. S. 1998, \mnras, 300, 945 (CHWF98)

\bibitem[Cole et al.\ 1999]{cole99}
Cole, S., Lacey, C., Baugh, C. \& Frenk, C.S. 1999, \mnras submitted

\bibitem[Couchman, H.M.P. 1991]{couchman91}
Couchman, H. M. P. 1991, \apj, 368, L23

\bibitem[Croft \& Gazta\~{n}aga 1997]{piza97}
Croft, R.A.C., \& Gazta\~{n}aga, E. 1997, \mnras, 285, 793

\bibitem[Davis \& Geller 1976]{davis76} 
Davis, M., \& Geller, M. J. 1976, \apj, 208, 13

\bibitem[Davis \& Peebles 1983]{dp83} 
Davis, M., \& Peebles, P.J.E. 1983, \apj, 267, 465

\bibitem[Davis et al. 1996]{davis96} 
Davis, M., Nusser, A., \& Willick, J.A. 1996, \apj, 473, 22

\bibitem[Dekel \& Lahav 1999]{dekel99}
Dekel, A. \& Lahav, O. 1999, \apj, 520, 24

\bibitem[de Lapparent, Geller \& Huchra 1986]{delapparent86}
de Lapparent, V., Geller, M.J. \& Huchra, J.P.  1986, \apj, L1 

\bibitem[Dressler 1980]{dressler80}
Dressler, A. 1980, \apj 236, 351

\bibitem[Efstathiou et al. 1992]{ebw92}
Efstathiou, G., Bond, J.R. \& White, S.D.M, 1992, \mnras, 258, 1P

\bibitem[Einasto et al.\ 1994]{einasto94}
Einasto, J., Saar, E., Einasto, M., Freudling, W., \& Gramann, M. 1994, \apj, 429, 465

\bibitem[Eke et al. 1996]{eke96} 
Eke, V.R, Cole, S., \& Frenk, C.S. 1996, \mnras, 282, 263

\bibitem[Fisher et al.\ 1994]{fisher94}
Fisher, K.B., Davis, M., Strauss, M.A., Yahil, A., \& Huchra, J. 1994, \mnras, 266, 50

\bibitem[Fisher et al.\ 1995]{fisher95}
Fisher, K.B., Huchra, J., Davis, M., Strauss, M.A., Yahil, A., \& Schlegel, D. 1995, \apjs, 100, 69

\bibitem[Freudling et al.\ 1999]{freudling99}
Freudling, W., et al. 1999, \apj, \apj, 523, 1

\bibitem[Gazta\~{n}aga et al. 1995]{gaztanaga95}
Gazta\~{n}aga, E., Croft, R.A.C., \& Dalton, G.B 1995, \mnras 276, 336

\bibitem[Gazta\~{n}aga \& Baugh 1998]{gaztanaga98}
Gazta\~{n}aga, E., \& Baugh, C.M 1998, \mnras, 294, 229


\bibitem[Giovanelli, Haynes \& Chincarini 1986]{giovanelli86}
Giovanelli, R.,  Haynes, M. P., \& Chincarini, G. L. 1986, \apj, 300, 77

\bibitem[Giovanelli \& Haynes 1989]{giovanelli89}
Giovanelli, R., \& Haynes, M.P., 1989, \aj, 97, 3

\bibitem[{Giovanelli et al. }1997]{giovanelli97}
Giovanelli, R., Haynes, M.P., Herter, T., Vogt, N.P., Wegner, G., Salzer, J.J., da Costa, L.N., \& Freudling, W. 1997, \aj, 113, 22

\bibitem[Giraud 1986]{giraud86}
Giraud, E.  1986, \aap, 170, 1

\bibitem[{Gramann }1993a]{gr93a}
Gramann, M. 1993a, \apj, 405, 449

\bibitem[{Gramann }1993b]{gr93b}
Gramann, M. 1993b, \apj, 405, L47

\bibitem[{Gramann et al. }1994]{grcg94}
Gramann, M., Cen, R., \& Gott, R.J. 1994, \apj, 425, 382

\bibitem[Guth \& Pi 1982]{guth82}
Guth, A.H., \& Pi, S.Y. 1982, Phys Rev Let, 49, 1110

\bibitem[Guzzo et al. 1997]{guzzo97}
Guzzo, L., Strauss, M. A., Fisher, K. B., Giovanelli, R. \& Haynes, M. P. 1997, \apj, 489, 37

\bibitem[Hamilton 1993]{hamilton93}
Hamilton, A.J.S. 1993, \apj, 417, 19

\bibitem[Hamilton 1998]{hamilton98}
Hamilton, A.J.S. 1998, in The Evolving Universe, ed. D. Hamilton, Kluwer Academic, p.\ 185

\bibitem[{Hawking }1982]{hawking82}
Hawking, S.W. 1982, Physics Letters, 115, 295

\bibitem[Hermit et al. 1996]{hermit96}
Hermit, S., Santiago, B.X., Lahav, O., Strauss, M.A., Davis, M., Dressler, A. \& Huchra, J. 1996, \mnras, 283, 709

\bibitem[Hockney \& Eastwood 1981]{hockney81}
Hockney, R. W., \& Eastwood, J. W. 1981, 
Computer Simulations Using Particles, (New York: McGraw Hill)

\bibitem[Hoffman \& Ribak 1991]{hoffman91}
Hoffman, Y. \& Ribak, E., 1991, \apj, 380, L5

\bibitem[Hubble 1936]{hubble36}
Hubble, E.P. 1936, The Realm of the Nebulae (Oxford University Press: Oxford), 79

\bibitem[Huchra \& Geller 1982]{huchra82}
Huchra, J. P., \& Geller, M. J. 1982, \apj, 257, 423

\bibitem[Hudson 1993]{hudson93}
Hudson, M.J. 1993, \mnras, 265, 43

\bibitem[Jenkins et al. 1998]{jenkins98} 
Jenkins, A., et al. 1998, \apj, 499, 20

\bibitem[{Kaiser }1987]{kaiser87}
Kaiser, N. 1987, \mnras, 227, 1

\bibitem[Kolatt et al.\ 1996]{kolatt96}
Kolatt, T., Dekel, A., Ganon, G., \& Willick, J.A. 1996, \apj, 458, 419

\bibitem[Lahav et al. 1990]{lahav90} 
Lahav, O., Nemiroff, R.J., \& Piran, Y. 1990, \apj, 350, 119

\bibitem[Lahav \& Saslaw 1992]{lahav92} 
Lahav. O., \& Saslaw, W.C. 1992, \apj, 396, 430

\bibitem[Lawrence et al. 1986]{lawrence86} 
Lawrence, A., Walker, D., Rowan-Robinson, M., Leech, K.J., \& Penston, M.V. 1986, \mnras, 219, 687

\bibitem[Little et al. 1991]{lwp91} 
Little, B., Weinberg, D.H, \& Park, C. 1991, \mnras, 253, 295

\bibitem[Little \& Weinberg 1994]{little94} 
Little, B., \& Weinberg, D.H. 1994, \mnras, 267, 605

\bibitem[Loveday et al. 1995]{loveday95} 
Loveday, J., Maddox, S. J., Efstathiou, G., \& Peterson, B. A. 1995, \apj, 468, 442

\bibitem[Lynden-Bell et al. 1989]{lyndenbell89}
Lynden-Bell, D., Lahav, O., \& Burstein, D. 1989, \mnras, 241, 325

\bibitem[Maddox et al.\ 1990]{maddox90} 
Maddox, S.J.M., Efstathiou, G., Sutherland, W.J, \& Loveday, J. 1990, \mnras, 242,43

\bibitem[Maddox et al.\ 1996]{maddox96} 
Maddox, S.J.M., Efstathiou, G., \& Sutherland, W.J 1996, \mnras, 283, 1227

\bibitem[{Meiksin \&  Davis} 1986]{meiksin86}
Meiksin, A., \&  Davis, M. 1986, \aj, 91, 191

\bibitem[{Melott \& Shandarin} 1990]{melott90}
Melott, A.L., \& Shandarin, S.F 1990, Nature, 346, 633

\bibitem[Melott \& Fry 1986]{melott86}
Melott, A.L., \&  Fry, J.N. 1986, \apj, 305, 1

\bibitem[{Melott \& Dominik} 1993]{md93}
Melott, A.L., \& Dominik, K.G. 1993, \apjs, 86, 1

\bibitem[Moore et al. 1994]{moore94}
Moore, B., Frenk, C.S., Efstathiou, G., \& Saunders, W. 1994, \mnras, 269, 742

\bibitem[Moore et al. 1993]{mfw93}
Moore, B., Frenk, C.S., \& White, S.D.M. 1993, \mnras, 261, 827

\bibitem[Moscardini et al. 1991]{moscardini91}
Moscardini, L., Matarrese, S., Lucchin, F. \& Messina, A. 1991, \mnras, 248, 424

\bibitem[{Narayanan \& Weinberg} 1998]{nw98}
Narayanan, V.K, \& Weinberg, D.H 1998, \apj, 508, 440 (NW98)

\bibitem[{Narayanan \& Croft} 1999]{nc99}
Narayanan, V.K, \& Croft, R.A.C. 1999, \apj, 515, 471

\bibitem[{Nolthenius \& White} 1987]{nw87}
Nolthenius, R., \& White, S.D.M 1987, \mnras, 235, 505

\bibitem[{Nusser \& Dekel} 1992]{nd92}
Nusser, A., \& Dekel, A. 1992, \apj, 391, 443

\bibitem[{Nusser, Dekel, \& Yahil} 1995]{ndy95}
Nusser, A., Dekel, A., \& Yahil, A. 1995, \apj, 449, 439

\bibitem[{Park, C.} 1990]{park90}
Park, C. 1990, PhD Thesis, Princeton University.

\bibitem[{Peacock \& Dodds} 1994]{pd94}
Peacock, J.A., \& Dodds, S.J. 1994, \mnras, 267, 1020

\bibitem[{Peebles }1980]{lss80}
Peebles, P.J.E. 1980, The Large Scale Structure of the Universe, (Princeton: Princeton Univ. Press)

\bibitem[Postman \& Geller 1984]{postman84}
Postman, M., \& Geller, M.J. 1984, \apj, 281, 95 (PG84)

\bibitem[ {Ryden \& Gramann} 1991]{rg91}
Ryden, B., \& Gramann, M. 1991, \apj, 383, L33

\bibitem[{Sandage }1986]{sandage86}
Sandage, A. 1986, \apj, 301, 1

\bibitem[{Sandage \& Tammann }1975]{sandage75}
Sandage, A., \& Tammann, G.A. 1975, \apj, 196, 313

\bibitem[Santiago et al.\ 1995]{ors1}
Santiago, B.X., Strauss, M.A., Lahav, O., Davis, M., Dressler, A., \& 

\bibitem[Santiago et al.\ 1996]{ors2}
Santiago, B.X., Strauss, M.A., Lahav, O., Davis, M., Dressler, A., \&

\bibitem[Sargent \& Turner 1977]{sargent77}
Sargent, W. W., \& Turner, E. L. 1977, ApJ, 212, L3

\bibitem[Saunders et al. 1992]{saunders92}
Saunders W., Rowan-Robinson, M., \& Lawrence, A. 1992, \mnras 258, 134

\bibitem[Saunders et al.\ 1995]{pscz1}
Saunders W., Sutherland, W., Efstathiou, G., \& Tadros, H. 1995, in Wide Field Spectroscopy and the Distant Universe, eds. S.J.Maddox \& Arag\'on-Salamanca, A., (Singapore: World Scientific), p.\ 88

\bibitem[Schlegel et al.\ 1994]{schlegel94}
Schlegel, D., Davis, M., Summers, F. \& Holtzman, J.A 1994, 427, 527

\bibitem[Sheth 1996]{sheth96}
Sheth, R.K. \mnras, 1996, 278, 101

\bibitem[Sigad et al.\ 1998]{sigad98}
Sigad, Y., Eldar, A., Dekel, A., Strauss, M. A., \& Yahil, A. 1998, \apj, 495, 516S

\bibitem[Smoot et al.\ 1991]{smoot91}
Smoot, G.F et al.\ 1991, \apj, 371, L1

\bibitem[Smoot et al.\ 1992]{smoot92}
Smoot, G.F et al.\ 1992, \apj, 396, L1

\bibitem[{Soda \& Suto} 1992]{soda92}
Soda, J., \& Suto, Y. 1992, \apj, 396, 379

\bibitem[Soifer et al. 1984]{soifer84}
Soifer, B.T. et al. 1984, \apj, 278, L71

\bibitem[{Springel \& White} 1998]{springel98}
Springel, V., White, S.D.M. 1998, \mnras, 298, 143

\bibitem[Starobinsky 1982]{starobinsky82}
Starobinsky, A. A. 1982, Phys Lett B, 117, 175

\bibitem[Strauss et al. 1992]{strauss92}
Strauss, M.A., Davis, M., Yahil, A., \& Huchra, J.P. 1992, \apj, 385, 421

\bibitem[Strauss \& Willick 1995]{strauss95}
Strauss, M.A. \& Willick, J.A., 1995, Physics Reports, 261, 271

\bibitem[Tadros et al. 1998]{tadros98}
Tadros, H., Efstathiou, G., \& Dalton, G.B 1998, \mnras, 296, 995

\bibitem[van de Weygaert \& Bertschinger 1996]{vandeweygaert96}
van de Weygaert, R., \& Bertschinger, E., 1996, \mnras, 281, 84V

\bibitem[{Viana \& Liddle} 1996]{viana96}
Viana, P.T.P, \& Liddle, A.R. 1996, \mnras, 281, 323

\bibitem[Vogeley et al. 1992]{vogeley92}
Vogeley, M.S., Park, C., Geller, M.J., \& Huchra, J.P 1992, \apj, 391, L5

\bibitem[{Weinberg } 1989]{weinberg89}
Weinberg, D.H. 1989, PhD Thesis, Princeton University.

\bibitem[{Weinberg }1992]{dhw92}
Weinberg, D.H. 1992, \mnras, 254, 315 

\bibitem[Weinberg \& Cole 1992]{wc92}
Weinberg, D.H., \& Cole, S. 1992, \mnras, 259, 652

\bibitem[{Weinberg, Gott \& Melott} 1987]{wgm87}
Weinberg, D.H.,  Gott, R.J., \& Melott, A.L., 1987, \apj, 321, 2

\bibitem[White 1979]{white79} 
White, S.D.M. 1979, \mnras, 186, 145

\bibitem[White et al. 1993]{wef93} 
White, S.D.M., Efstathiou, G., \& Frenk, C.S. 1993, \mnras, 262, 1023

\bibitem[Whitmore, Gilmore, \& Jones 1993]{whitmore93}
Whitmore, B. C., Gilmore, D. M., \& Jones, C. 1993, \apj, 407, 489

\bibitem[Willick et al.\ 1997a]{willick97a}
Willick, J.A., Strauss, M.A., Dekel, A., \& Kolatt, T. 1997a, \apj, 486, 629

\bibitem[Willick et al.\ 1997b]{willick97b}
Willick, J.A., Courteau, S., Faber, S.M., Burstein, D., Dekel, A., \& Strauss, M.A. 1997b, \apjs, 109, 333

\bibitem[Willick \& Strauss 1998]{willick98}
Willick, J. A., \& Strauss, M. A. 1998, \apj, 507, 64

\bibitem[Willmer et al. 1998]{willmer98}
Willmer, C. N. A., da Costa, L. N., \& Pellegrini, P. S. 1998, \aj, 115, 869

\bibitem[Yahil et al.\ 1991]{yahil91}
Yahil, A., Strauss, M.A., Davis, M. \& Huchra, J.P., 1991, \apj, 372, 380

\bibitem[{Zel'dovich }1970]{zel70}
Zel'dovich, Ya. B. 1970 A\&A, 5, 84

\bibitem[Zwicky 1937]{zwicky37}
Zwicky, F. 1937, \apj, 86, 217

\end{thebibliography}
\end{document}